\documentclass[ECO,biber]{nowfnt} 

\usepackage[utf8]{inputenc}
\usepackage{amsmath}
\usepackage{mathtools}
\usepackage{adjustbox}
\usepackage{booktabs}
\usepackage{subfigure}
\usepackage{pdflscape}

\newcommand{\BEAS}{\begin{eqnarray*}}
   \newcommand{\EEAS}{\end{eqnarray*}}
   \newcommand{\BEA}{\begin{eqnarray}}
   \newcommand{\EEA}{\end{eqnarray}}
   \newcommand{\BEQ}{\begin{equation}}
   \newcommand{\EEQ}{\end{equation}}
   \newcommand{\BIT}{\begin{itemize}}
   \newcommand{\EIT}{\end{itemize}}
   \newcommand{\BNUM}{\begin{enumerate}}
   \newcommand{\ENUM}{\end{enumerate}}

   \newcommand{\BA}{\begin{array}}
   \newcommand{\EA}{\end{array}}
   
   \newcommand{\eg}{{\it e.g.}}
   \newcommand{\ie}{{\it i.e.}}

   \newcommand{\ones}{\mathbf 1}
   
   \newcommand{\reals}{{\mbox{\bf R}}}

   \newcommand{\symm}{{\mbox{\bf S}}}  
   
   \newcommand{\Tr}{\mathop{\bf Tr}}
   \newcommand{\diag}{\mathop{\bf diag}}

   \newcommand{\Expect}{\mathop{\bf E{}}}

   \newcommand{\Cov}{\mathop{\bf cov{}}}

   \newcommand{\argmax}{\mathop{\rm argmax}}

   {\begin{quote}}{\end{quote}}
   
   \newcounter{oursection}

   \newcounter{lecture}

\newcommand{\sigmahat}{\hat{\Sigma}}

\newcommand{\lthat}{\hat{L}^T}
\newcommand{\lhat}{\hat{L}}

\newcommand{\rtilde}{\tilde{r}}

\title{A Simple Method for\\ Predicting Covariance Matrices\\ of Financial Returns}

\maintitleauthorlist{
Kasper Johansson \\
Stanford University \\
kasperjo@stanford.edu
\and
Mehmet G. Ogut \\
Stanford University \\
giray98@stanford.edu
\and 
Markus Pelger \\
Stanford University \\
mpelger@stanford.edu
\and
Thomas Schmelzer \\
Stanford University \\
Abu Dhabi Investment Authority \\
thomas.schmelzer@adia.ae
\and
Stephen Boyd \\
Stanford University \\
boyd@stanford.edu
}

\issuesetup
{%
copyrightowner={A.~Heezemans and M.~Casey},
volume        = xx,
issue         = xx,
pubyear       = 2023,
isbn          = xxx-x-xxxxx-xxx-x,
eisbn         = xxx-x-xxxxx-xxx-x,
doi           = 10.1561/XXXXXXXXX,
firstpage     = 1, 
lastpage      = 87
}

\usepackage{mwe}

\author[1]{Johansson, Kasper}
\author[2]{Ogut, Mehmet G.}
\author[3]{Pelger, Markus}
\author[4]{Schmelzer, Thomas}
\author[5]{Boyd, Stephen}

\affil[1]{Department of Electrical Engineering, Stanford University; kasperjo@stanford.edu}
\affil[2]{Department of Electrical Engineering, Stanford University;
giray98@stanford.edu}
\affil[3]{Department of Management Science and Engineering, Stanford University;
mpelger@stanford.edu}
\affil[4]{Department of Electrical Engineering, Stanford University, and Abu Dhabi Investment Authority;
thomas.schmelzer@adia.ae}
\affil[5]{Department of Electrical Engineering, Stanford University;
boyd@stanford.edu}

\articledatabox{\nowfntstandardcitation}

\begin{document}

\makeabstracttitle

\clearpage 

\begin{abstract}
We consider the well-studied problem of predicting the time-varying covariance
matrix of a vector of financial returns. Popular methods range from simple
predictors like rolling window or exponentially weighted moving average (EWMA)
to more sophisticated predictors such as generalized autoregressive conditional
heteroscedastic (GARCH) type methods.  Building on a specific covariance
estimator suggested by Engle in 2002, we propose a relatively simple extension
that requires little or no tuning or fitting, is interpretable, and produces
results at least as good as MGARCH, a popular extension of GARCH that handles
multiple assets. To evaluate predictors we introduce a novel approach,
evaluating the regret of the log-likelihood over a time period such as a
quarter. This metric allows us to see not only how well a covariance predictor
does over all, but also how quickly it reacts to changes in market conditions.
Our simple predictor outperforms MGARCH in terms of regret. We also
test covariance predictors on downstream applications such as portfolio
optimization methods that depend on the covariance matrix. For these
applications our simple covariance predictor and MGARCH perform similarly.
\end{abstract}

\chapter{Introduction}
\label{c-intro} 
\section{Covariance prediction} \label{s-cov-pred}
We consider
cross-sections, \eg, a vector time series of $n$ financial returns,
denoted $r_t\in \reals^n$, $t=1,2,\ldots$, where $(r_t)_i$ is the return of 
asset $i$ from $t-1$ to $t$. 
We focus on the case where the mean $\Expect r_t$ is small enough that
the second moment $\Expect r_tr_t^T\in \reals^{n \times n}$
is a good approximation of the covariance matrix
$\Cov(r_t) = \Expect r_tr_t^T - (\Expect r_t)(\Expect r_t)^T$, where $\Expect$
denotes expectation. This is the case
for most daily, weekly, or monthly stock, bond, and futures returns, 
factor returns, and index returns.  
We start by focussing on the case where the number of assets $n$ is modest,
say, on the order 10--100 or so; in
chapter~\ref{c-large-universes} we explain how to extend the method
to much larger universes using ideas such as factor models.

We model the returns $r_t$ as independent random variables with zero mean and
covariance $\Sigma_t \in \symm_{++}^n$ (the set of symmetric positive
definite matrices). We focus on the problem of predicting or estimating
$\Sigma_t$, based on knowledge of $r_1, \ldots, r_{t-1}$. The prediction is
denoted as $\sigmahat_t \in \symm_{++}^n$. 
The predicted volatilities of assets are given by
\[
\hat\sigma_t = \diag (\sigmahat_t)^{1/2}\in \reals^n,
\]
where $\diag$ with a matrix argument is the
vector of diagonal entries of the matrix, and the squareroot of a vector above 
is elementwise. 
We denote the predicted correlations as
\[
\hat R_t = \diag(\hat\sigma_t)^{-1} \sigmahat_t \diag(\hat\sigma_t)^{-1},
\]
where $\diag$ with a vector argument is the diagonal matrix with entries 
from the vector argument.

Covariance estimation comes up in several areas of finance, including Markowitz
portfolio construction~\citep{markowitz_1952,grinold2000_portfolio}, risk
management~\citep{mcneil2015quantitative}, and asset pricing~\citep{sharpe_1964}.
Much attention has been devoted to this problem, and a Nobel Memorial Prize in
Economic Sciences was awarded for work directly related to volatility
estimation~\citep{engle_1982}.

While it is well known that the tails of financial returns are poorly modeled by
a Gaussian distribution, our focus here is on the bulk of the distribution,
where the Gaussian assumption is reasonable. For future use, we note that the
log-likelihood of an observed return $r_t$, under the Gaussian distribution $r_t
\sim \mathcal N(0,\sigmahat_t)$, is 
\BEQ\label{e-ll}
l_t(\sigmahat_t) =
\frac{1}{2}\left(-n\log(2\pi) - \log\det \sigmahat_t - r_t^T \sigmahat_t^{-1}
r_t\right). 
\EEQ
The Gaussian log-likelihood is closely related to a popular metric
for evaluating covariance predictors in econometrics,
called the (Gaussian) quasi-likelihood 
(QLIKE)~\citep{patton2011volatility, patton2009evaluating, laurent2013loss}.
QLIKE is the negative log-likelihood, under the Gaussian assumption,
up to an additive constant and a positive scale factor.
Roughly speaking, we seek covariance predictors that achieve
large values of log-likelihood, or small values of QLIKE,
on realized returns.  We will describe evaluation
of covariance predictors in detail in chapter~\ref{c-eval}.

\section{Contributions}
This monograph makes three contributions. First, we propose a new method for
predicting the time-varying covariance matrix of a vector of financial returns,
building on a specific covariance estimator suggested by Engle in
2002.  Our method is a relatively simple extension that requires very little
tuning and is readily interpretable.  
It relies on solving a small convex optimization
problem, which can be carried out very quickly and reliably \citep{boyd2004convex}.
Our method performs as well as much more complex methods, as measured by several metrics.

Our second contribution is to propose a new method for evaluating a covariance
predictor, by considering the regret of the log-likelihood over some time period
such as a quarter. This approach allows us to evaluate how quickly a
covariance estimator reacts to changes in market conditions.

Our third contribution is an extensive empirical study of covariance predictors.
We compare our new method to other popular predictors, including rolling window,
exponentially weighted moving average (EWMA),
and generalized autoregressive conditional
heteroscedastic (GARCH) type methods.  We find that our method performs slightly better
than other predictors. However, even the simplest predictors perform well for
practical problems like portfolio optimization.

Everything needed to reproduce our results, together with an open
source implementation of our proposed covariance predictor, is available 
online at
\begin{quote}
\centering
\url{https://github.com/cvxgrp/cov_pred_finance}.
\end{quote}

\section{Outline}
In chapter~\ref{c-common-predictors} we
describe some common predictors, including the one that our method builds on. 
We introduce our proposed covariance predictor in chapter~\ref{c-wite}.
In chapter~\ref{c-eval} we
discuss methods for validating covariance predictors that measure both overall
performance and reactivity to market changes.
We describe the data we use in our first empirical studies
in chapter~\ref{c-exp}, and give the results in chapter~\ref{c-results}.

In the next chapters we discuss some extensions of and variations on our method,
including realized covariance prediction (chapter~\ref{c-realized-cov}),
handling large universes via factor models (chapter~\ref{c-large-universes}),
obtaining smooth covariance estimates (chapter~\ref{c-smooth}),
and using our covariance model to generate simulated returns
(chapter~\ref{c-generative}).

\chapter{Some common covariance predictors}\label{c-common-predictors} In this
chapter we review some common covariance predictors, ranging from simple to
complex, with the goal of giving context and fixing our notation. To simplify
some formulas, we take $r_\tau=0$ for $\tau \leq 0$.

\section{Rolling window}
The rolling window predictor with window length or memory $M$ is the average of
the last $M \geq n$ outer products,
\[ 
\sigmahat_t = \alpha_t\sum_{\tau=t-M}^{t-1} r_\tau r_\tau^T, \quad t=2,3, \ldots,
\]
where $\alpha_t = 1/\min\{t-1,M\}$ is the normalization constant. The rolling
window predictor can be evaluated via the recursion 
\[ 
\sigmahat_{t+1} = \frac{\alpha_{t+1}}{\alpha_{t}}\sigmahat_{t} + \alpha_{t+1}(r_t r_t^T-r_{t-M}r_{t-M}^T),
\quad t=1,2, \ldots,
\]
with initialization $\sigmahat_1=0$. 

For $t<n$, the rolling window covariance estimate is not full rank. To handle
this, as well as to improve the quality of the prediction, we can add
regularization or shrinkage, for example by adding a positive multiple of
$\diag(\sigmahat_t)$ to our estimate \citep{ledoit110_honey_cov,
ledoit2003improved}, or approximating the predicted covariance matrix
by a diagonal plus low rank matrix, as described in chapter~\ref{c-large-universes}.

\section{EWMA} \label{s-ewma}
The exponentially weighted moving average (EWMA) estimator, with forgetting
factor $\beta \in(0,1)$, is
\BEQ\label{e-ewma}
\sigmahat_t = \alpha_{t} \sum_{\tau=1}^{t-1}\beta^{t-1-\tau} r_\tau r_\tau^T,
\quad t=2,3, \ldots,
\EEQ
where 
\[
\alpha_t=\left( \sum_{\tau=1}^{t-1}\beta^{t-1-\tau} \right)^{-1} = \frac{1-\beta}{1-\beta^{t-1}}
\]
is the normalization constant. The forgetting factor $\beta$ is usually
expressed in terms of the half-life $H=-\log 2 / \log\beta$, for which $\beta^H
=1/2$. The half-life $H$ is the number of periods when the exponential weight has
decreased by a factor of two. For example, for a
half-life of one year, the current observed return has twice the impact on our
covariance prediction as the return observed one year ago. The EWMA predictor is
widely used in practice; for example RiskMetrics 
suggests the forgetting factor $\beta=0.94$, which
corresponds to a half-life of around 11 days~\citep{menchero2011barra,
longerstaey1996riskmetrics}.

The EWMA covariance predictor can be computed recursively as
\[ 
\sigmahat_{t+1} =  \frac{\beta-\beta^{t}}{1-\beta^t}
\sigmahat_t + \frac{1-\beta}{1-\beta^t}r_tr_t^T, \quad t=1,2, \ldots,
\]
with initialization $\sigmahat_1=0$.
Like the rolling window predictor, the EWMA predictor is singular for $t<n$,
which can be handled using the same regularization methods described above.

\section{GARCH and MGARCH}
\paragraph{GARCH.}
The generalized autoregressive conditional heteroscedastic (GARCH) predictor
decomposes the return of a single asset as
\[
r_t = \mu + \epsilon_t,  
\]
where $\mu$ is the mean return and $\epsilon_t$ is the innovation, and models the
innovation as
\[
\epsilon_t = \sigma_t z_t,  \qquad 
\sigma_t^2 = \omega +  \sum_{\tau=1}^q a_{\tau} \epsilon_{t-\tau}^2  + \sum_{\tau=1}^p b_{\tau} \sigma_{t-\tau}^2,
\]
where $\sigma_t$ is the asset volatility, $z_t$ are independent $\mathcal N(0,1)$, and $q$ and $p$ (often both set to one in
practice) determine the GARCH order~\citep{bollerslev_1986}. (Recall that we assume zero mean.)  
The model parameters are $\omega$, $a_1,
\ldots, a_q$, and $b_1, \ldots, b_p$. Estimating the model
parameters requires solving a nonconvex optimization
problem~\citep{cov_barrat_2022}.

With $p=0$ we recover the autoregressive conditional heteroscedastic
(ARCH) predictor, introduced in the seminal paper by~\citet{engle_1982}. This paper set the foundation for a
wide variety of popular volatility and correlation predictors and earned him the
2003 Nobel Memorial Prize in Economic Sciences. 

\paragraph{MGARCH.}
There are several ways of extending the GARCH predictor to a multivariate or
vector setting. The most popular is the dynamic conditional correlation
(DCC) predictor~\citep{DCC}, which is a two-step approach described below.

Many other MGARCH predictors have been proposed. The most
straightforward generalization from the univariate to multivariate predictors is the 
VEC predictor, where the covariance matrix
is vectorized and each element is modeled as a GARCH process with dependencies
on all other elements~\citep{bollerslev_engle_1988}. However, this extension
requires estimating $n(n+1)(n(n+1)+1)/2 \approx n^4/2$
parameters, which can be impractical even for modest values of $n$.

Following the VEC extension of GARCH, multivariate GARCH (MGARCH)
predictors have been proposed in two lines of
development~\citep{silvennoinen2009multivariate}. The first line involves models
that impose restrictions on the parameters of the VEC predictor, including
DVEC~\citep{bollerslev_1986}, BEKK~\citep{engle_kroner_1995},
FF-MGARCH~\citep{FF_MGARCH}, O-GARCH~\citep{OGARCH}, and GO-GARCH~\citep{GOGARCH},
to name some. However, these predictors have been shown to be hard to fit and
can yield inconsistent estimates~\citep{brooks_2003}. (These inconsistencies may
not have much practical impact.)
For detailed reviews of MGARCH predictors we refer the reader
to~\citep{silvennoinen2009multivariate, garch_survey}

\section{DCC GARCH}

The second line of extensions of GARCH to vector time series 
models conditional covariances
through separate estimates of conditional variances and
correlations~\citep{DCC, engle2001theoretical}. 
In \citep{CCC} Bollerslev introduced the constant conditional correlation 
predictor (CCC) where the
individual asset volatilities are modeled as separate GARCH processes, while the
correlation matrix is assumed constant and equal to the unconditional
correlation matrix. This predictor was later extended to the dynamic conditional
correlation (DCC) predictor where the correlation matrix is allowed to change
over time~\citep{DCC}.
The DCC model has the form 
\[
\Sigma_t = D_tR_tD_t,
\]
where $D_t$ is the diagonal matrix of standard deviations, \ie,
$(D_t)_{ii}=(\Sigma_t)_{ii}^{1/2}$,
and $R_t$ is the
correlation matrix associated with $\Sigma_t$. 

DCC GARCH models the
diagonal elements of $D_t$ as separate univariate GARCH processes as described above. The
correlation matrix $R_t$ is then modeled as a constrained multivariate GARCH (MGARCH)
process, \eg, as
\BEAS
R_t &=& \diag(\diag(Q_t))^{-1/2} Q_t\diag(\diag(Q_t))^{-1/2}, \\ 
Q_t &=& \bar{Q} (1-a-b) + a \rtilde_t \rtilde_t^T + b Q_{t-1},
\EEAS
where $\bar{Q}$ is the unconditional correlation matrix, $a$ and $b$ are the
MGARCH parameters, and $\rtilde_t$ are the volatility adjusted returns defined
as 
\[
\rtilde_t = D_t^{-1} r_t.  
\]
The parameters can be estimated in two steps via (quasi) maximum likelihood, but 
requires solving non-convex optimization problems~\citep{DCC}.
This predictor has become a
popular choice amongst MGARCH predictors due to its interpretability. 
Variants of the DCC predictor are widely used in finance,
where it is also often used in combination with EWMA estimates. Conditional
correlation predictors are easier to estimate than other multivariate GARCH
predictors, and their parameters are more interpretable. 

\paragraph{Iterated covariance estimation.}
DCC, which separately estimates the volatilities and correlations, is closely
related to the idea of iterated covariance predictors \citep{cov_barrat_2022}. 
Iterated covariance predictors estimate the covariance matrix in multiple
iterations.  In a two-step iteration we first form a first covariance estimate
$\sigmahat_t^{(1)}$ of the returns $r_t$, at each time $t$, and form the whitened returns
\[
\rtilde_t = \left(\sigmahat_t^{(1)}\right)^{-1/2} r_t. 
\]
In the second iteration we form the covariance estimate $\sigmahat_t^{(2)}$ of
the whitened returns $\rtilde_t$. The final covariance estimate (of the returns
$r_t$) is then formed as
\[
\sigmahat_t = \left(\sigmahat_t^{(1)}\right)^{1/2} \sigmahat_t^{(2)} \left(\sigmahat_t^{(1)}\right)^{1/2}.
\]
This procedure can be iterated further, and has been shown empirically to
improve the quality of the covariance estimate; see~\citep{cov_barrat_2022} for
details.
In DCC, $\hat \Sigma^{(1)}$ is diagonal and models the volatilities;
$\hat \Sigma^{(2)}$ is a correlation matrix.

\section{Iterated EWMA}\label{s-iewma} Iterated EWMA (IEWMA) was proposed
by~\citep{DCC} and is analogous to DCC GARCH but with EWMA estimates of the
volatilities and correlations instead of GARCH. Engle proposed IEWMA as an
efficient alternative to the DCC GARCH predictor, although he did not refer to
it as IEWMA; we use this term to emphasize its connection to iterated whitening,
as proposed in~\citep{cov_barrat_2022}. Specifically, IEWMA can be viewed as an
iterated whitener, where we first use a diagonal whitener (which estimates the 
volatilities) and then a full matrix whitener (which estimates the
correlations). This is analogous to the two-step iterated covariance predictor
where $\sigmahat_t^{(1)}$ is the diagonal matrix of squared volatility estimates and
$\sigmahat_t^{(2)}$ estimates the correlation matrix of the volatility adjusted
returns.

First we form an estimate of the volatilities 
$\hat\sigma_t=\diag(\sigmahat_t)^{1/2}$ using EWMA
predictors for each asset.
We denote the half-life of these volatility estimates as $H^\text{vol}$.
We then form the marginally standardized returns as 
\BEQ \label{e-tilde-rt}
\tilde r_t = \hat D_t^{-1} r_t,
\EEQ
where $\hat D_t = \diag(\hat \sigma_t)$. These vectors should have entries with
standard deviation near one. It is common practice to winsorize the standardized
returns; a good rule of thumb is to clip $\tilde r_t$ at $\pm 4.2$, which corresponds
to clipping $r_t$ at $\pm 4.2 \hat \sigma_t$.

Then we form a EWMA estimate of the covariance of $\tilde r_t$, which we denote
as $\tilde R_t$, using half-life $H^\text{cor}$ for this EWMA estimate. (We use
the superscript `cor' since the diagonal entries of $\tilde R_t$ should be near
one, so $\tilde R_t$ is close to a correlation matrix.) From $\tilde R_t$ we
form its associated correlation matrix $\hat R_t$, \ie, we scale $\tilde R_t$ 
on the left and right by a diagonal matrix with entries 
$(\tilde R_t)_{ii}^{-1/2}$. Since the diagonal
entries of $\tilde R_t$ should be near one, $\tilde R_t$ and $\hat R_t$ are not too
different.

Our IEWMA covariance predictor is
\[
\sigmahat_t = \hat D_t \hat R_t \hat D_t, \quad t=2,3, \ldots.
\]
This is the covariance predictor proposed in \citep{DCC}; replacing
$\hat R_t$ with $\tilde R_t$ we obtain the iterated whitener proposed by Barratt and
Boyd in \citep{cov_barrat_2022}. As mentioned above, they are typically quite
close.

It is common to choose the volatility half-life $H^\text{vol}$ to be smaller
than the correlation half-life $H^\text{cor}$. The intuition here is that we can
average over fewer past samples when we predict the $n$ volatilities
$\hat{\sigma}_t$, but need more past samples to reliably estimate the $n(n-1)/2$
off-diagonal entries of $\hat R_t$. Empirical studies on real return data confirm
that choosing a faster volatility half-life than correlation half-life yields
better estimates.

\chapter{Combined multiple iterated EWMAs}\label{c-wite} In this chapter we
introduce a novel covariance predictor, which we call combined multiple iterated
EWMAs, for which we use the acronym CM-IEWMA. The CM-IEWMA predictor is
constructed from a modest number of IEWMA predictors, with different pairs of
half-lives, which are combined using dynamically varying weights that are based
on recent performance.

The CM-IEWMA predictor is motivated by the idea that different pairs of
half-lives may work better for different market conditions. For example, short
half-lives perform better in volatile markets, while long half-lives
perform better for calm markets where conditions are changing slowly. 

\section{Dynamically weighted prediction combiner} \label{s-dynamic-combiner}
We first describe the idea in a general setting. We start with $K$ different
covariance predictors, denoted $\hat \Sigma^{(k)}_t$, $k=1, \ldots, K$. These
could be any of the predictors described above, or predictors of the same type
with different parameter values, \eg, half-lives (for EWMA) or pairs of
half-lives (for IEWMA). In some contexts these different predictors are
referred to as a set of $K$ experts~\citep{hastie2009elements,
jordan1994hierarchical}. 

We denote the Cholesky factorizations of the associated precision matrices
$(\sigmahat_t^{(k)})^{-1}$ as $\hat L_t^{(k)}$, \ie,
\[
\left(\sigmahat_t^{(k)}\right)^{-1} = \hat L_t^{(k)} 
( \hat L_t^{(k)} )^T, \quad k=1, \ldots, K,
\]
where $\hat L_t^{(k)}$ are lower triangular with positive diagonal entries. We will combine these Cholesky factors with nonnegative weights $\pi_1,\ldots, \pi_K$
that sum to one, to obtain
\BEQ\label{e-Lhat}
\hat L_t = \sum_{k=1}^K \pi_k \hat L_t^{(k)}.
\EEQ
From this we recover the weighted combined predictor
\BEQ\label{e-combined}
\hat \Sigma_t = \left( \hat L_t \hat L_t^T \right)^{-1}.
\EEQ
We will see below why we combine the Cholesky factors of the precision 
matrices, and not the covariance or precision matrices themselves.

\section{Choosing the weights via convex optimization}
The log-likelihood \eqref{e-ll} can be expressed in terms of the Cholesky factor
of the precision matrix $\hat L_t$ as
\[
l_t (\sigmahat_t) 
= -(n/2)\log(2\pi) + \sum_{i=1}^n \log \lhat_{t,ii} - (1/2)\|\lthat_tr_t\|^2_2,
\]
where $\| \cdot \|_2$ denotes the Euclidean norm.
This is a concave function of the weights $\pi\in
\reals_+^K$~\citep{boyd2004convex}.

We choose the weights at time $t$
as the solution of the convex optimization problem
\BEQ\label{e-w-it-prob}
\begin{array}{ll}
\mbox{maximize} 
& 
\sum_{j=1}^N\bigg(\sum_{i=1}^n \log \lhat_{t-j,ii} -
(1/2)\|\lthat_{t-j}r_{t-j}\|^2_2\bigg) \\
\mbox{subject to}  &\lhat_\tau = \sum_{j=1}^{K}\pi_j\hat L_\tau^{(j)}, \quad
\tau=t-1,\dots,t-N \\
\quad & \pi \geq 0, \quad \ones^T\pi=1, 
\end{array}
\EEQ
with variables $\pi_1, \ldots, \pi_K$, where $N$ is the look-back, $\ones$
denotes the vector with entries one, and $\geq$ between vectors means entrywise.
In words: we choose the (mixture) weights in each period so as to maximize the average
log-likelihood of the combined prediction over the trailing $N$ periods. The
problem \eqref{e-w-it-prob} is convex, and can be solved very quickly and reliably
by many methods \citep{boyd2004convex}.
The covariance predictor is then recovered using \eqref{e-Lhat} and \eqref{e-combined}.

The look-back $N$ is a parameter that can be adjusted to give good performance.
Numerical experiments suggest that the predictor is not very sensitive to the 
choice of $N$, and that a choice $N=10$ seems to work well for 
asset universes up to a few hundred assets.

We mention several extensions of the weight problem \eqref{e-w-it-prob}. First,
we can add one prediction which is diagonal, using any estimates of the
volatilities (including constant). This gives us shrinkage, automatically
chosen. We can also add a constraint or objective term
that encourages the weights to vary smoothly over time, as discussed more
in chapter~\ref{c-smooth}.

The CM-IEWMA predictor is a special case of the dynamically weighted prediction
combiner described above, where the $K$ predictions are each IEWMA, with
different pairs of half-lives $H^\text{vol}$ and $H^\text{cor}$.

\chapter{Evaluating covariance predictors} \label{c-eval}
There are several ways of evaluating a covariance predictor, often divided into
two categories, direct and indirect~\citep{patton2009evaluating},~\citep[\S
7]{ANDERSEN2006777}. Direct methods use a proxy for the true covariance matrix
to evaluate the predictor, while indirect methods use the covariance predictor
on tasks of interest, such as portfolio construction or portfolio tracking. 

Popular direct methods are the 
Mincer-Zarnowitz (MZ) regression and its
variants, based on statistical tests of
the regression coefficients of a predicted variable on an observed variable (or
in the case of variance and covariance, a proxy for the observed
variable)~\citep{mincer1969evaluation, theil1961economic}. Direct methods also
include the comparison between different predictors in terms of some 
loss function. Common loss functions are the mean squared error (MSE)
and quasi-likelihood (QLIKE)~\citep{patton2011volatility, patton2009evaluating}.
To select good models, the model confidence set (MCS) is usually
used~\citep{hansen2011model}, or the Ledoit--Wolf test 
\citep{ledoit2008robust} to compare Sharpe ratios. 

Indirect methods use applications to rank covariance
predictors, and include the minimum variance and
mean-variance portfolios, as well as portfolio tracking tasks. 

The
difference in performance between various predictors can also be evaluated using
statistical tests. For a more detailed discussion of both direct and indirect
methods, we refer the reader to~\citep{patton2009evaluating}.

In this chapter we discuss several evaluation metrics for covariance predictors. 
The first three metrics are direct, and include the mean squared error and two
metrics based on a statistical measure, 
the log-likelihood under a Gaussian distribution.  The
remaining metrics judge a covariance predictor by the performance of a portfolio
using a method that depends on a covariance matrix. We are mainly interested in
illustrating how simple methods can perform just as well as or better than more
complex ones, rather than finding optimal predictors in a statistical sense.
Therefore we look at the absolute performance of covariance predictors on these
metrics.

\section{Mean squared error}
The mean squared error (MSE) is a common metric for evaluating a covariance
predictor $\sigmahat_t$, defined as
\[
\frac{1}{T} \sum_{t=1}^T \|r_tr_t^T - \sigmahat_t\|_F^2,
\]
\ie, the average squared Frobenius norm of the difference between the realized
(rank one) covariance matrix $r_tr_t^T$ and the covariance predictor $\sigmahat_t$.
Lower values of MSE are better. 
One variation on the MSE error assumes that $\sigmahat_t$ is constant over some
number of time periods and replaces the rank one realized covariance $r_tr_t^T$ 
with an average of the rank one terms over the periods, \ie, the realized
empirical covariance.

\section{Log-likelihood}
A natural way of judging a covariance predictor is via its average
log-likelihood on realized returns,
\[
\frac{1}{2T}\sum_{t=1}^T\Big( -n\log(2\pi) - \log\det \sigmahat_t - r_t^T \sigmahat_t^{-1}r_t\Big),
\]
with larger values being better. This metric can be used to compare different
predictors.

To understand the performance of a covariance predictor over time and
changing market conditions, we can examine the average log-likelihood over
periods such as quarters, and look at the distribution of quarterly
average log-likelihood values. We are particularly interested in poor, \ie, low
values.

\section{Log-likelihood regret}
Recall that the best constant predictor, in terms of the log-likelihood, is the
empirical sample covariance
\[
\Sigma^\text{emp} = \frac{1}{T} \sum_{t=1}^T  r_t r_t^T ,
\]
with value
\[
\frac{1}{2}\big(-n(\log(2\pi)+1) - \log\det \Sigma^\text{emp}\big).
\]
For any other constant $\Sigma \in \symm_{++}^n$, the log-likelihood is lower
than the log-likelihood of $\Sigma^\text{emp}$. We define the \emph{average
log-likelihood regret} as the average log-likelihood of the (constant) empirical
covariance, minus the average log-likelihood of the covariance predictor. The
regret is a measure of how much the covariance predictor $\sigmahat_t$, $t=1,
\ldots, T$, underperforms the best possible constant covariance predictor (\ie,
the sample covariance matrix). The term regret comes from the field of online
optimization; see, \eg, \citep{zinkevich2003_regret, Mokhtari2016_regret,
hazan2007_regret, hazan2016introduction}. 

We want our covariance predictor to have small regret. The regret is typically
positive, but it can be negative, \ie, our time-varying covariance can have
higher log-likelihood than the best constant one. The regret is not any more
useful than the log-likelihood when comparing predictors over one time interval,
since it simply adds a constant and switches the sign. But it is interesting
when we compute the regret over multiple periods, like months or quarters. The
regret over multiple quarters removes the effect of the log-likelihood of the
empirical covariance varying due to changing market conditions, and allows us to
assess how well the covariance predictor adapts.

\section{Portfolio performance}\label{s-pp} We can also judge the performance
of a covariance predictor by the investment performance of portfolio
construction methods that depend on the estimated covariance matrix. As with
log-likelihood or log-likelihood regret, we can examine the portfolio
performance in periods such as quarters, to see how evenly the performance is
spread over time. 

One obvious metric of interest is how close the ex-ante and realized
portfolio volatilities are. The metrics described above, MSE,
log-likelihood, and log-likelihood regret, are agnostic
to the portfolio; with specific real portfolios we can see how well our
covariance predictors predict portfolio volatility.

We will assess a covariance predictor using five simple portfolio construction
methods.  
The first is an equally weighted (or $1/n$) portfolio, which does not by itself
depend on the covariance, but does when we adjust it with cash to achieve a
given ex-ante risk. The second, third, and fourth portfolios depend only on the
covariance matrix. They are minimum variance, risk parity, and maximum
diversification portfolios. For an in depth discussion of these portfolios,
see~\citep{braga2015risk}. The last portfolio we consider is a mean-variance
portfolio, using a very simple mean estimator.

For each portfolio we look at four metrics: realized return, volatility, Sharpe
ratio, and maximum drawdown. The returns, volatilities, and
Sharpe ratios are reported in annualized values. The Sharpe ratio is defined as
the ratio of the excess return (over the risk-free rate), divided by the
volatility of the excess return,
\[
\frac{\frac{1}{T}\sum_{\tau=1}^T 
(r^p_t-r^{\text{rf}}_t)}{\Big(\frac{1}{T}\sum_{\tau=1}^T
\big( r^p_t - \frac{1}{T}\sum_{\tau=1}^T r^p_t\big)^2\Big)^{1/2}},
\]
where $r^p_t$ and $r^{\text{rf}}_t$ are the portfolio and risk-free returns at
time $t$. 
The maximum drawdown is defined as
\[
\max_{1 \leq t_1 < t_2\leq T} \frac{V^p_{t_1}}{V^p_{t_2}} - 1,
\]
where
\[
V^p_t = V_0 (1+r^p_1)(1+r^p_2) \cdots (1+r^p_t)
\]
is the portfolio value at time $t$ (with returns re-invested), starting with
value $V_0>0$.

In addition to portfolio performance, we can also examine how well the
covariance prediction predicts the portfolio volatility.   We compare the
realized or ex-post portfolio volatility
\[
\left( \frac{1}{T} \sum_{t=1}^T (r_t^Tw_t)^2 \right)^{1/2},
\]
to the predicted or ex-ante portfolio volatility
\[
\left( \frac{1}{T} \sum_{t=1}^T w_t^T \sigmahat_t w_t\right)^{1/2},
\]
where $w_t\in\reals^n$ are the portfolio weights.
This directly measures
the ability of the estimated covariance matrix to predict portfolio risk.

\paragraph{Equal weight portfolio.}
We take the equal weight or $1/n$ portfolio with $w= (1/n)\ones$. This portfolio
does not depend on the covariance $\sigmahat_t$, but when we mix it with cash,
as described below, it will.

\paragraph{Minimum variance portfolio.}
The (constrained) minimum variance portfolio is the solution of the convex
optimization problem
\[
\begin{array}{ll}
\mbox{minimize} &  \quad w^T\sigmahat_t w \\
\mbox{subject to} & \quad w^T\ones = 1, \quad \|w\|_1 \leq L_{\max}, \quad w_{\min} \leq w \leq w_{\max}
\end{array}
\]
with variable $w\in \reals^n$, where $L_{\max} \geq 1$ is a leverage limit, and $w_{\min}$
and $w_{\max}$ are lower and upper bounds on the weights, respectively. 

\paragraph{Risk-parity portfolio.}
The portfolio return volatility $\sigma(w) = (w^T \sigmahat_t w)^{1/2}$ can be
broken down into a sum of volatilities (risks) associated with each asset as
\[
\frac{\partial \log \sigma(w)}{\partial w_i} = 
\frac{\partial \sigma(w)}{\sigma(w)} \frac{w_i}{\partial w_i} =\frac{w_i(\sigmahat_t w)_i}
{w^T \sigmahat_t w}, \quad i=1,\dots,n.
\]
The risk parity portfolio is the one for which these volatility attributions are
equal~\citep{Qian119}. This portfolio can be found by solving the convex
optimization problem~\citep{cvx_book_additional},
\[
\mbox{minimize} \quad (1/2)x^T\sigmahat_t x - \sum_{i=1}^n (1/n)\log x_i,
\]
with variable $x$, and then taking $w=x^{\star}/(\ones^Tx^{\star})$.

\paragraph{Maximum diversification portfolio.}
The diversification ratio of a long-only portfolio (\ie, one with $w \geq 0$) is
defined as
\[
D(w) =\frac{\hat \sigma_t^T w}{(w^T\sigmahat_t w)^{1/2}}.   
\]
The diversification ratio tells us how much higher the portfolio volatility
would be if all assets were perfectly correlated. The maximum diversification
portfolio is the portfolio $w$ that maximizes $D(w)$, possibly subject to
constraints~\citep{choueifaty2008toward}.
Like the risk-parity portfolio, the maximum diversification portfolio can be
found via convex optimization. We let $x^\star$ denote the solution of the
convex optimization problem 
\citep{cvx_book_additional}
\[
\begin{array}{ll}
\mbox{minimize} &  x^T \sigmahat_t x\\
\mbox{subject to} & \hat \sigma_t^T x = 1, \quad x \geq 0, 
\end{array}
\]
with variable $x$. The maximum diversification portfolio is
$w=x^{\star}/\ones^T x^{\star}$.

\paragraph{Volatility control with cash.}
We mix each of the four portfolios described above with cash to achieve a target
value of ex-ante volatility $\sigma^\mathrm{tar}$. To do this we start with the
portfolio weight vector $w_t$, and compute its ex-ante volatility $\sigma_t =
(w_t^T \sigmahat_t w_t)^{1/2}$. Then we add a cash component so that the overall
ex-ante volatility equals our target, \ie, we use the $(n+1)$ weights (with the
last component denoting cash)
\[
\left[ \begin{array}{c}
\theta w_t \\ (1-\theta)
\end{array}\right], \qquad \theta = \frac{\sigma^\text{tar}}{\sigma_t}.
\]
This portfolio will have ex-ante volatility $\sigma^\mathrm{tar}$. Note that the
cash weight can be either positive (when it dilutes the portfolio volatility) or
negative (when it leverages the portfolio volatility to the desired level). The
target volatility $\sigma^\text{tar}$ should be chosen so as to avoid portfolios
that are either too diluted or too leveraged.

\paragraph{Mean variance portfolio.}
The last portfolio we consider is a basic mean-variance portfolio, defined as
the solution of the convex optimization problem
\[
\begin{array}{ll}
\mbox{maximize} &  \quad  \hat{r}_t^T w \\
\mbox{subject to} & \quad \|\sigmahat_t^{1/2} w\|_2 \leq \sigma^\text{tar} \\
& \quad \ones^T w + c = 1, \quad \|w\|_1 \leq L_{\max}, \\ & \quad w_{\min} \leq w \leq w_{\max}, \quad c_{\min} \leq c \leq c_{\max}
\end{array}
\]
with variable $w$, where $\hat{r}_t$ is the predicted mean return vector at time $t$.
The vector $w$ gives the weights of the non-cash assets and $c$ denotes the cash
weight.
The non-cash and cash weights are limited by $w_{\min}, w_{\max}$ and
$c_{\min}, c_{\max}$, respectively. This portfolio
does not need cash dilution, since it includes cash in its construction. (If
$\sigma^\text{tar}$ is chosen appropriately, it will have ex-ante risk
$\sigma^\text{tar}$.) The mean-variance portfolio depends not only on a
covariance estimate, but also a return estimate.  For this we use one of the
simplest possible return estimates, a EWMA of the realized returns.

\chapter{Data sets and experimental setup}\label{c-exp} We illustrate our
method on three different data sets: a set of $49$ industry portfolios, a set
of $25$ stocks, and a set of $5$ factor returns, each augmented with cash (with
the historical risk-free interest rate). For each data set we show results for
six covariance predictors. Everything needed to reproduce the results is
available online at
\begin{quote}
\centering
\url{https://github.com/cvxgrp/cov_pred_finance}.
\end{quote}

\section{Data sets}\label{s-data}
\paragraph{Industry portfolios.}
The first data set consists of the daily returns of a universe of $n=49$ daily
traded industry portfolios, shown in table~\ref{t-industries}, along with cash.
The data set spans July 1st 1969 to December 30th, 2022, for a total of 13496
(trading) days. The data was obtained from the Kenneth French Data
Library~\citep{french_data_lib}.
\begin{table} \footnotesize
\centering
\caption{Industry portfolios.}
\label{t-industries}
\begin{adjustbox}{center}
\begin{tabular}{ll}
\toprule
Agriculture & Food products \\
Candy \& soda & Beer \& liquor \\
Tobacco products & Recreation \\
Entertainment & Printing and publishing \\
Consumer goods & Apparel \\
Healthcare & Medical equipment \\
Pharmaceutical products & Chemicals \\
Rubber and plastic products & Textiles \\
Construction materials & Construction \\
Steel works etc. & Fabricated products \\
Machinery & Electrical equipment \\
Automobiles and trucks & Aircraft \\
Shipbuilding, railroad equipment & Defense \\
Precious metals & Non-metallic and industrial metal mining \\
Coal & Petroleum and natural gas \\
Utilities & Communication \\
Personal services & Business services \\
Computers & Computer software \\
Electronic equipment & Measuring and control equipment \\
Business supplies & Shipping containers \\
Transportation & Wholesale \\
Retail & Restaurants, hotels, motels \\
Banking & Insurance \\
Real estate & Trading \\
Other & \\
\bottomrule
\end{tabular}
\end{adjustbox}
\end{table}

\paragraph{Stocks.}
The second data set consists of the daily returns of $n=25$ stocks and cash. The
stocks were chosen to be the 25 largest stocks in the S\&P 500 at the beginning
of 2010, listed in table~\ref{t-stocks}. This data set spans January 4th 2010 to
December 30th, 2022, for a total of 3272 (trading) days. The stock data was
attained through the Wharton Research Data Services (WRDS) portal~\citep{WRDS}.
\begin{table} \small
\centering
\caption{List of companies and their tickers.}
\label{t-stocks}
\begin{tabular}{ll}
\toprule
Ticker & Company Name \\
\midrule
XOM   & Exxon Mobil \\
WMT   & Walmart \\
AAPL  & Apple Inc. \\
PG    & Procter \& Gamble \\
JNJ   & Johnson \& Johnson \\
CHL   & China Mobile \\
IBM   & IBM \\
SBC   & AT\&T \\
GE    & General Electric \\
CHV   & Chevron \\
PFE   & Pfizer \\
NOB   & Noble \\
NCB   & NCR \\
KO    & Coca-Cola \\
ORCL  & Oracle Corporation \\
HWP   & Hewlett-Packard \\
INTC  & Intel Corporation \\
MRK   & Merck \& Co. \\
PEP   & PepsiCo \\
BEL   & Becton, Dickinson and Company \\
ABT   & Abbott Laboratories \\
SLB   & Schlumberger \\
P     & Pandora Media \\
PA    & Pan American Silver \\
MCD   & McDonald's \\
\bottomrule
\end{tabular}%
\label{tab:companies}%
\end{table} 
\paragraph{Factor returns.}
The third data set consists of daily returns of the five Fama-French factors
taken from the Kenneth French Data Library~\citep{french_data_lib}, shown in
table~\ref{t-factors}. The data set spans July 1st 1963 to December 30th, 2022,
for a total of 14979 (trading) days.
\begin{table}
    \centering
    \caption{The five Fama-French factors.}
    \label{t-factors}
    \begin{adjustbox}{center}\small
    \begin{tabular}{p{0.3\linewidth} p{0.6\linewidth}}
    \toprule
    \textbf{Factor} & \textbf{Description} \\
    \midrule
    MKT-Rf & market excess return over risk-free rate \\
    SMB & small stocks minus big stocks \\
    HML & high book-to-market stocks minus low book-to-market stocks \\
    RMW & stocks with high operating profitability minus stocks with low operating profitability \\
    CMA & stocks with conservative investment policies minus stocks with aggressive investment policies \\
    \bottomrule
    \end{tabular}
    \end{adjustbox}
    \end{table}  





\begin{figure}
\centering
\subfigure[Industry data set.]{\includegraphics[height=0.35\linewidth]{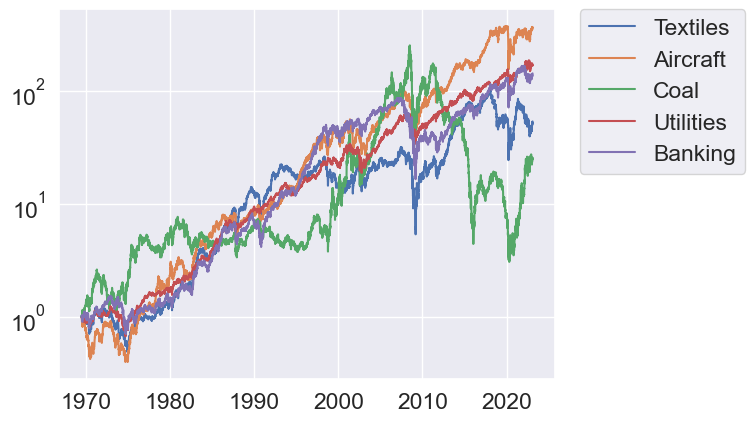}}\\
\subfigure[Stock data set.]{\includegraphics[height=0.35\linewidth]{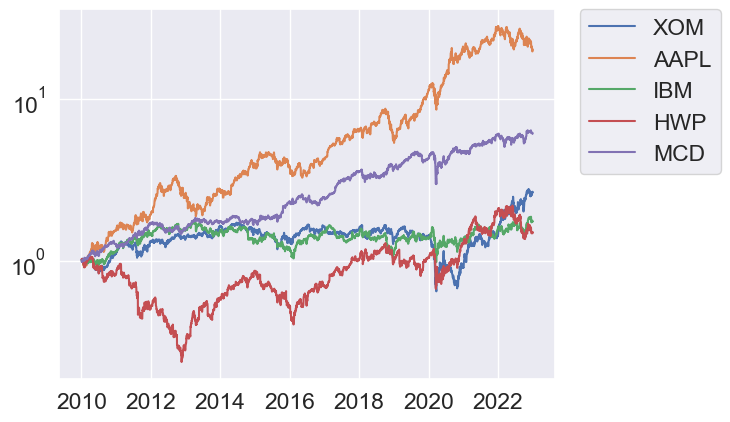}}\\
\subfigure[Factor data set.]{\includegraphics[height=0.35\linewidth]{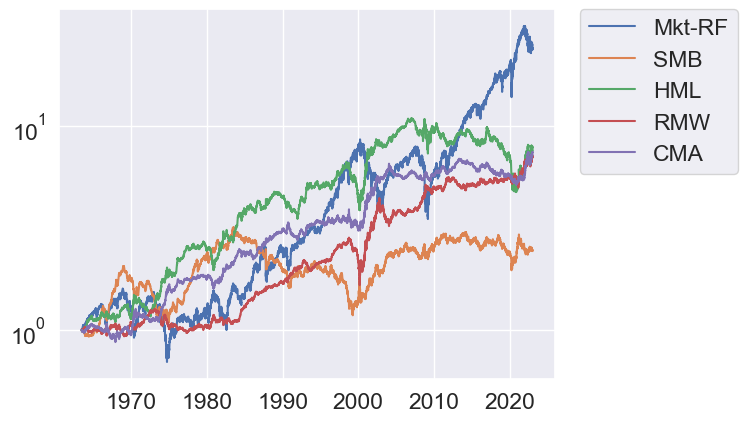}}
\caption{Cumulative returns of five assets from each data set.}
\label{f-cumsum-returns}
\end{figure}

\clearpage
\paragraph{Cumulative returns.}
In figure~\ref{f-cumsum-returns} we show the cumulative returns of
the five factors, and the cumulative returns of five assets chosen 
from each of the industry and stock data sets.

\clearpage
\section{Six covariance predictors}\label{s-experiment-predictors}
For each data set we evaluate six covariance predictors, described below. 
\BIT
\item Rolling window estimates with 500-, 250-, and, 125-day windows for the
industry, stock, and factor data sets, respectively, denoted RW in plots and tables.
\item EWMA predictors with 250-, 125-, and, 63-day half-lives, for the
industry, stock, and factor data sets, respectively, denoted EWMA.
\item IEWMA predictors with half-lives (in days) $H^\text{vol}/H^\text{cor}$ of 125/250,
63/125, and 21/63 for the three data sets, respectively, denoted IEWMA.
\item DCC GARCH predictor, denoted MGARCH, with parameters re-estimated annually
using the \texttt{rmgarch} package in \textsf{R}~\citep{ghalanos2019rmgarch}.
\item CM-IEWMA predictor with $K=5$ IEWMA predictors and a lookback of $N=10$
days, with half-lives shown in table~\ref{t-halflives}. For each of the fastest
IEWMA predictors we regularize the covariance estimate by increasing the
diagonal entries by 5\%. 
\item Prescient predictor, \ie, the empirical covariance for the quarter the day
is in. This predictor maximizes log-likelihood for each quarter, and achieves
zero regret. It is of course not implementable, and meant only to show a bound
on performance with which to compare our implementable predictors.
\EIT
\begin{table}
\centering
\caption{Half-lives for CM-IEWMA predictors, given as
$H^\text{vol}/H^\text{cor}$, in days.}
\label{t-halflives}
\begin{center}
\begin{tabular}{l|ccccc}
\toprule
\textbf{Data set} & \textbf{Half-lives} \\
\midrule
Industries & 21/63 & 63/125 & 125/250 & 250/500 & 500/1000 \\
Stocks &  
10/21 & 21/63 & 63/125 & 125/250 & 250/500\\
Factors & 5/10 & 10/21 & 21/63 & 63/125 & 125/250\\
\bottomrule
\end{tabular}
\end{center}
\end{table}

All the parameters above (\eg, half-lives) are chosen as reasonable values that
give good overall performance for each predictor. The results are not sensitive
to these choices.

For our experiments we use the first two years (500 data points) of each data
set to fit the MGARCH predictor and initialize the other predictors. (After
this initial MGARCH fit, we re-estimate its parameters annually.) Hence, the
evaluation period for our experiments below ranges from June 24th 1971 to
December 30th, 2022, for the industry portfolios, from December 28th, 2011, to
December 30th, 2022, for the stock portfolios, and from June 28th 1965 to December
30, 2022, for the factor portfolios.

\chapter{Results}\label{c-results}
\section{CM-IEWMA component weights}
\begin{figure}
\centering
\subfigure[Industry data
set.]{\includegraphics[width=0.6\linewidth]{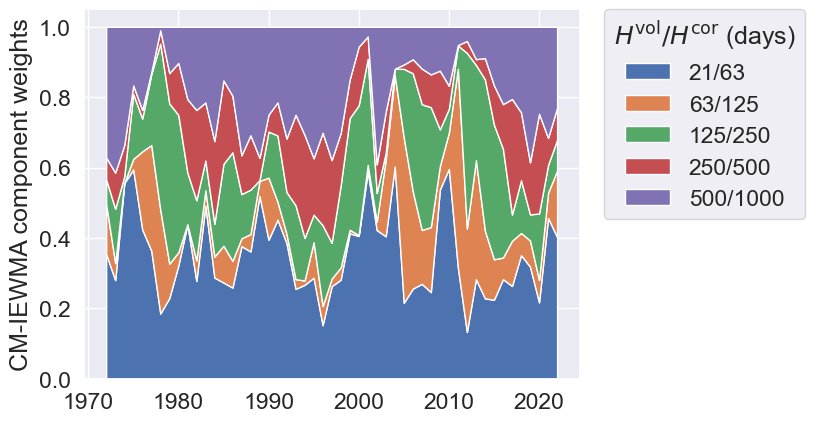}}\\
\subfigure[Stock data
set.]{\includegraphics[width=0.6\linewidth]{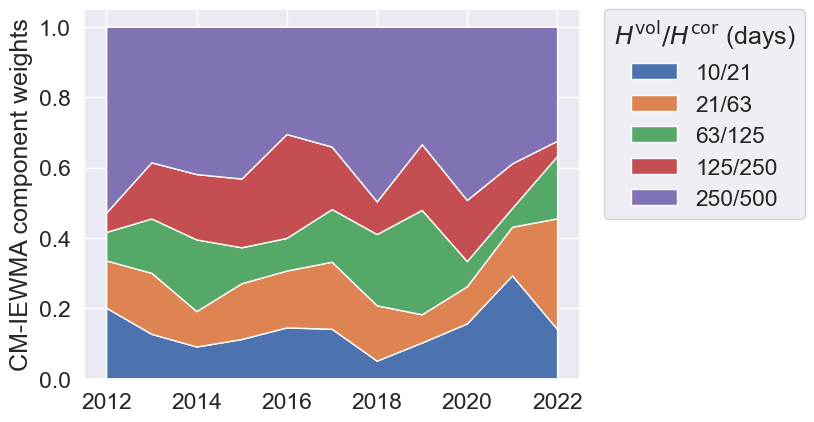}}\\
\subfigure[Factor data
set.]{\includegraphics[width=0.6\linewidth]{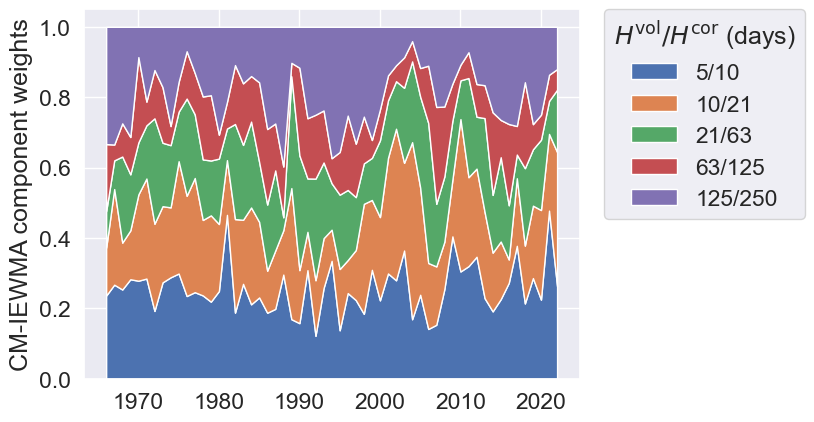}}
\caption{Weights of the various IEWMA components in the CM-IEWMA predictors on
three data sets. The
IEWMA components are
represented as $H^\text{vol}/H^\text{cor}$ for the volatility and correlation
half-lives, respectively.}
\label{f-ewma_weights}
\end{figure}

Figure~\ref{f-ewma_weights} shows the weights for each of the five
components of the CM-IEWMA predictors, averaged yearly, for the three
data sets.

\begin{figure}
\centering
\subfigure[Industry data
set.]{\includegraphics[width=0.5\linewidth]{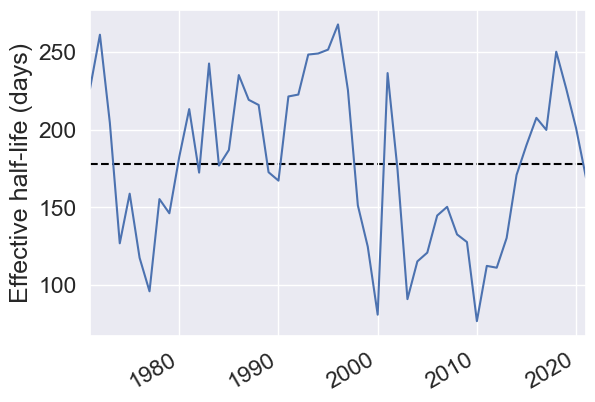}}\\
\subfigure[Stock data
set.]{\includegraphics[width=0.5\linewidth]{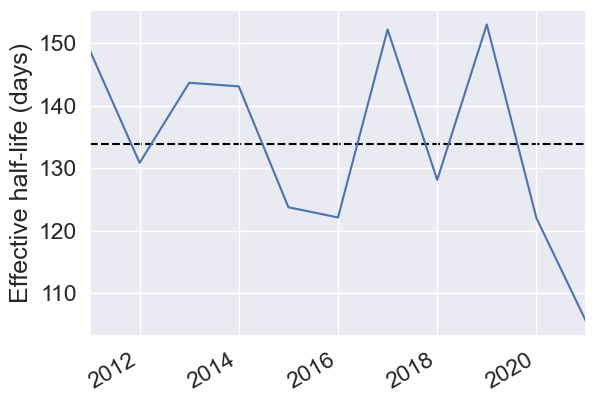}}\\
\subfigure[Factor data
set.]{\includegraphics[width=0.5\linewidth]{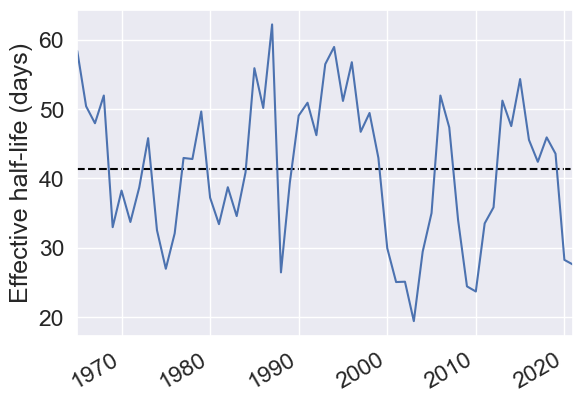}} 
\caption{Effective half-lives of the CM-IEWMA predictor on three data sets.}
\label{f-effective_T}
\end{figure}

We can see how the predictor adapts the weights depending on market conditions.
Substantial weight is put on the slower (longer half-life) IEWMAs most years.
During and following volatile periods like the 2000 dot.com bubble or 2008
market crash, we see a big increase in weight on the faster IEWMAs. We
can illustrate these changes in weights in response to market conditions via the
effective half-life of the CM-IEWMA, defined as the weighted average of the five
(longer) half-lives, shown in figure~\ref{f-effective_T}, averaged yearly.

\clearpage

\section{Mean squared error}
Table~\ref{t-mse} shows the average, standard deviation, and maximum of the MSE
computed over distinct quarters for the six covariance predictors on the three data sets
(with lower being better for all three metrics).
\begin{table}
\centering
\caption{Metrics on the MSE, computed over distinct quarters, for six covariance predictors on three data sets.}
\label{t-mse}
\begin{subtable}
\centering
\begin{tabular}{lccc}
\toprule
Predictor & Average/$10^{-4}$ & Std. Dev./$10^{-3}$ & Max/$10^{-2}$ \\
\midrule
RW & $7.6$ & $4.0 $ & $3.9$ \\
EWMA & $7.5$ & $4.0$ & $3.9$ \\
IEWMA & $7.4$ & $3.9$ & $3.9$ \\
MGARCH & $\mathbf{6.8}$ & $\mathbf{3.6}$ & $\mathbf{3.8}$ \\
CM-IEWMA & $6.9$ & $\mathbf{3.6}$ & $\mathbf{3.8}$ \\
\hline
Prescient & $6.6$ & $3.5$ & $3.7$ \\
\bottomrule
\end{tabular}
\captionof*{subtable}{Industry data set.}
\end{subtable}

\vspace{1em}

\begin{subtable}
\centering
\begin{tabular}{lccc}
  \toprule
  Predictor & Average/$10^{-7}$ & Std. Dev./$10^{-6}$ & Max/$10^{-5}$ \\
  \midrule
  RW & $3.4$ & $1.9$ & $2.4$ \\
  EWMA & $3.4$ & $1.9$ & $2.4$ \\
  IEWMA & $3.3$ & $1.8$ & $2.4$ \\
  MGARCH & $\mathbf{3.2}$ & $\mathbf{1.8}$ & $2.4$ \\
  CM-IEWMA & $\mathbf{3.2}$ & $\mathbf{1.8}$ & $2.4$ \\
  \hline
  Prescient & $3.1$ & $1.8$ & $2.3$ \\
  \bottomrule
  \end{tabular}

\captionof*{subtable}{Stock data set.}
\end{subtable}

\vspace{1em}

\begin{subtable}
\centering
\begin{tabular}{lccc}
\toprule
Predictor & Average/$10^{-4}$ & Std. Dev./$10^{-3}$ & Max/$10^{-2}$ \\
\midrule
RW & $3.4$ & $1.6$ & $1.1$ \\
EWMA & $3.3$ & $1.6$ & $1.1$ \\
IEWMA & $3.2$ & $1.6$ & $1.1$ \\
MGARCH & $3.0$ & $\mathbf{1.4}$ & $1.0$ \\
CM-IEWMA & $\mathbf{2.9}$ & $\mathbf{1.4}$ & $\mathbf{0.9}$ \\
\hline
Prescient & $3.0$ & $1.5$ & $1.0$ \\
\bottomrule
\end{tabular}

\captionof*{subtable}{Factor data set.}
\end{subtable}
\end{table}
CM-IEWMA and MGARCH do better than the other predictors on all metrics over all
data sets, with MGARCH doing slightly better on the industry data and CM-IEWMA
slightly better on the stock data. Interestingly, on the factor data set, the
CM-IEWMA predictor does better than the prescient predictor.

\section{Log-likelihood and log-likelihood regret}
\begin{figure}
\centering
\subfigure[Industry data
set.]{\includegraphics[width=0.6\textwidth]{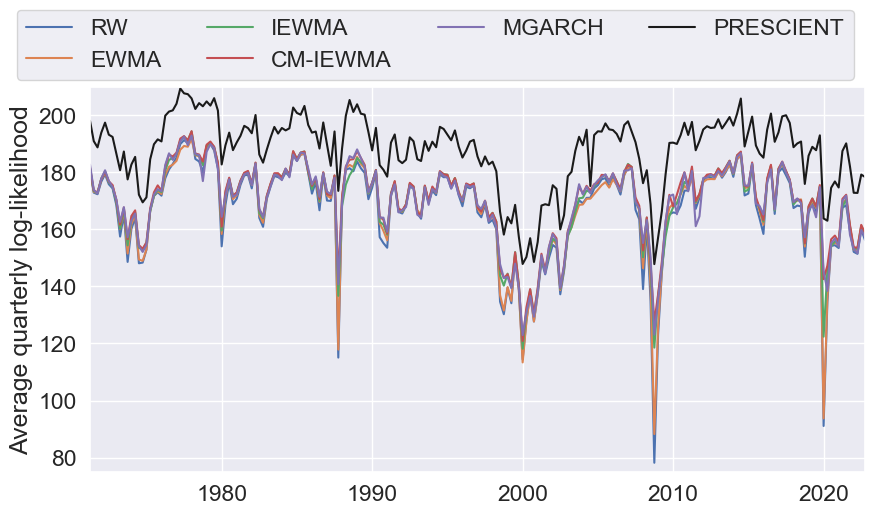}}\\
\subfigure[Stock data
set.]{\includegraphics[width=0.6\textwidth]{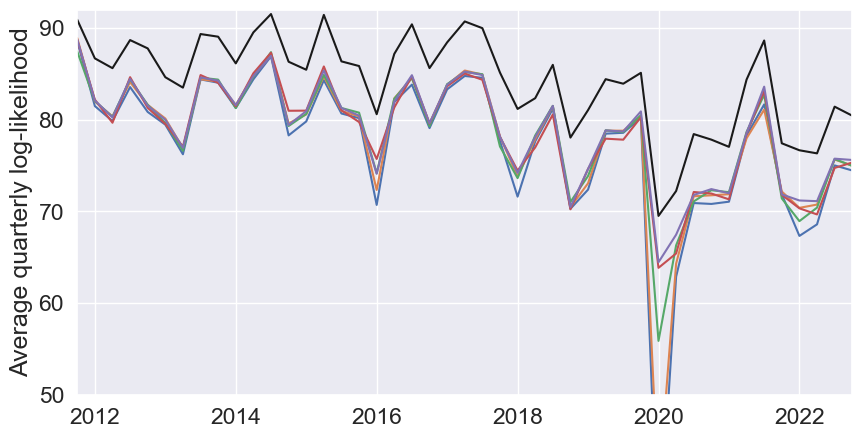}}\\
\subfigure[Factor data
set.]{\includegraphics[width=0.6\textwidth]{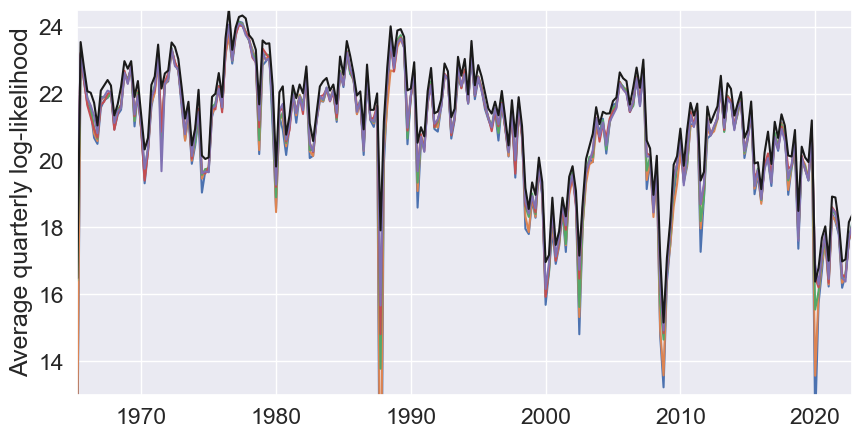}} 
\caption{The log-likelihood, averaged quarterly, for six covariance predictors and three data sets.}
\label{f-ll}
\end{figure}

Figure~\ref{f-ll} shows the average quarterly log-likelihood for the different
covariance predictors over the evaluation period. Not surprisingly, the
prescient predictor does substantially better than the others. The different
predictors follow similar trends, with even the prescient predictor experiencing
a drop in log-likelihood during market turbulence. Close inspection shows that
the CM-IEWMA and MGARCH predictors almost always have the highest log-likelihood
in each quarter.

\begin{figure}
\centering
\subfigure[Industry data
set.]{\includegraphics[width=0.6\textwidth]{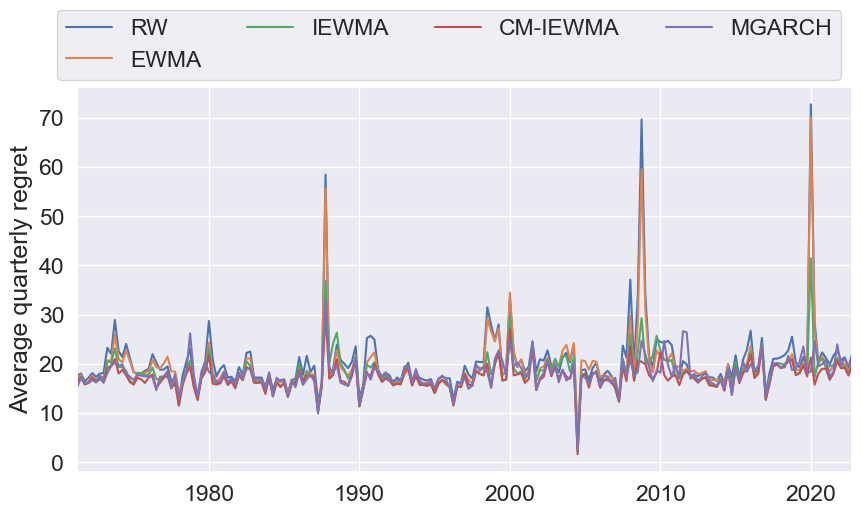}}\\
\subfigure[Stock data
set.]{\includegraphics[width=0.6\textwidth]{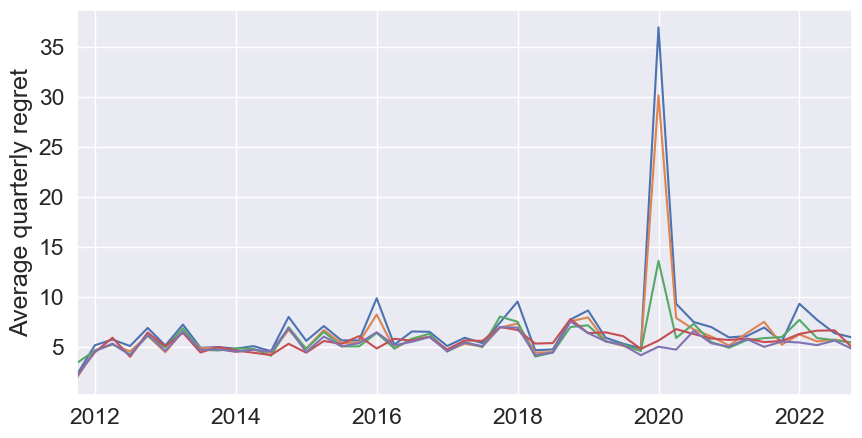}}\\
\subfigure[Factor data
set.]{\includegraphics[width=0.6\textwidth]{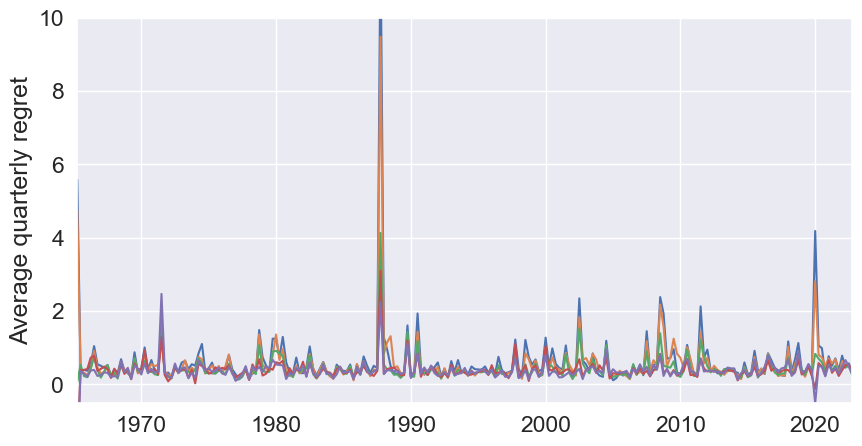}} 
\caption{The regret, averaged quarterly, for five covariance predictors over the evaluation periods for three data sets.}
\label{f-regret}
\end{figure}

\begin{figure}
\centering
\subfigure[Industry data
set.]{\includegraphics[width=0.45\textwidth]{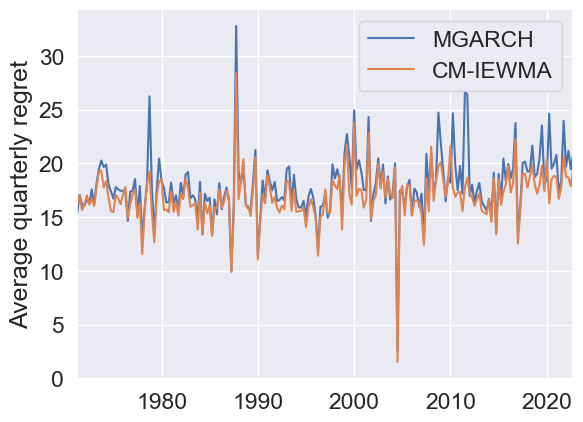}}\\
\subfigure[Stock data
set.]{\includegraphics[width=0.45\textwidth]{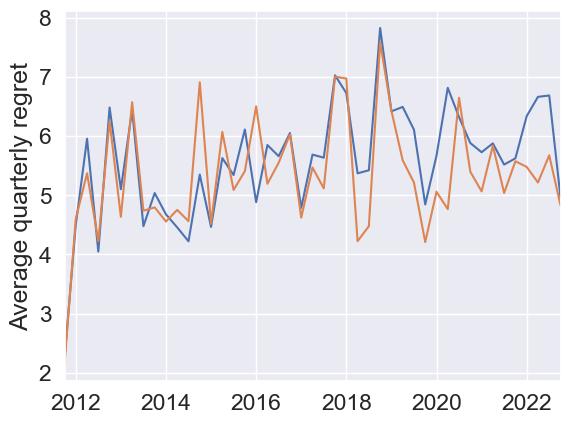}}\\
\subfigure[Factor data
set.]{\includegraphics[width=0.45\textwidth]{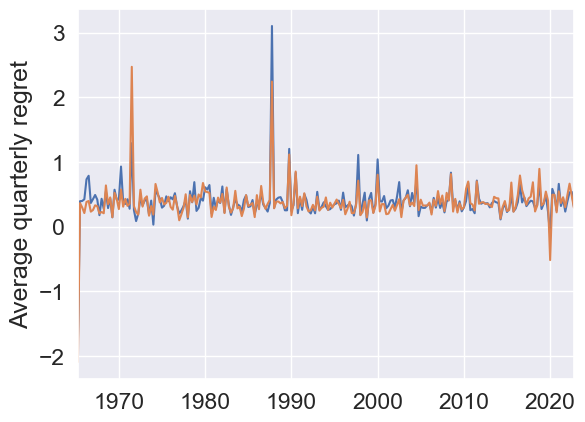}} 
\caption{The regret for MGARCH and CM-IEWMA, averaged quarterly over the evaluation periods for three data sets.}
\label{f-regret_ewma_vs_garch}
\end{figure}

Figure~\ref{f-regret} shows the average quarterly log-likelihood regret for the
different covariance predictors over the evaluation period. Clearly, CM-IEWMA
and MGARCH perform best in volatile markets. Figure~\ref{f-regret_ewma_vs_garch}
illustrates the difference between CM-IEWMA and MGARCH. As seen, CM-IEWMA
consistently has lower regret on the industry and stock data sets, while they
perform similar on the factor data. More precisely, CM-IEWMA has lower regret
than MGARCH in 87\% of the quarters for the industry data, 71\% for the stock
data, and 51\% for the factor data.

\begin{table}
\centering
\caption{Metrics on the average quarterly regret for six covariance predictors on three data sets.}
\label{t-regret}
\begin{subtable}
\centering
\begin{tabular}{lccccc}
\toprule
Predictor & Average & Std.~dev. & Max \\
\midrule
RW & 20.4 & 6.9 & 72.8 \\
EWMA & 19.4 & 6.2 & 70.1 \\
IEWMA & 18.2 & 3.6 & 41.4 \\
MGARCH & 17.9 & 3.0 & 32.8 \\
CM-IEWMA & \textbf{16.9} & \textbf{2.4} & \textbf{28.4} \\
\hline
PRESCIENT & 0.0 & 0.0 & 0.0 \\
\bottomrule
\end{tabular}
\captionof*{subtable}{Industry data set.}
\end{subtable}

\vspace{1em}

\begin{subtable}
\centering
\begin{tabular}{lccccc}
\toprule
Predictor & Average & Std.~dev. & Max \\
\midrule
RW & 7.0 & 4.8 & 37.0 \\
EWMA & 6.2 & 3.8 & 30.2 \\
IEWMA & 5.8 & 1.6 & 13.6 \\
MGARCH & 5.6 & \textbf{1.0} & 7.8 \\
CM-IEWMA & \textbf{5.3} & \textbf{1.0} & \textbf{7.6} \\
\hline
PRESCIENT & 0.0 & 0.0 & 0.0 \\
\bottomrule
\end{tabular}

\captionof*{subtable}{Stock data set.}
\end{subtable}

\vspace{1em}

\begin{subtable}
\centering
\begin{tabular}{lccccc}
\toprule
Predictor & Average & Std.~dev. & Max \\
\midrule
RW & 0.6 & 0.9 & 12.2 \\
EWMA & 0.6 & 0.7 & 9.5 \\
IEWMA & \textbf{0.4} & \textbf{0.3} & 4.1 \\
MGARCH & \textbf{0.4} & \textbf{0.3} & 3.1 \\
CM-IEWMA & \textbf{0.4} & \textbf{0.3} & \textbf{2.9} \\
\hline
PRESCIENT & 0.0 & 0.0 & 0.0 \\
\bottomrule
\end{tabular}

\captionof*{subtable}{Factor data set.}
\end{subtable}
\end{table}

Table~\ref{t-regret} illustrates the differences in regret further, by showing
the average, standard deviation, and the maximum
of the average quarterly regret.
As we can see, the average quarterly regret is lower for CM-IEWMA than for the
other predictors. The regret is also more stable for CM-IEWMA, as the standard
deviation is lower. Finally, the maximum average quarterly regret is also lower for CM-IEWMA than for the other predictors. These results are most prominent on the
industry and stock data, while MGARCH does similar on the factor data.

\begin{figure}
\centering
\subfigure[Industry
data set.]{\includegraphics[height=0.33\textwidth]{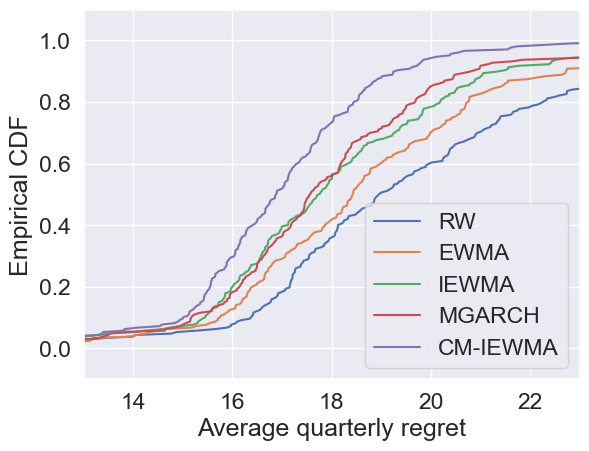}}
\\\subfigure[Stock data
set.]{\includegraphics[height=0.33\textwidth]{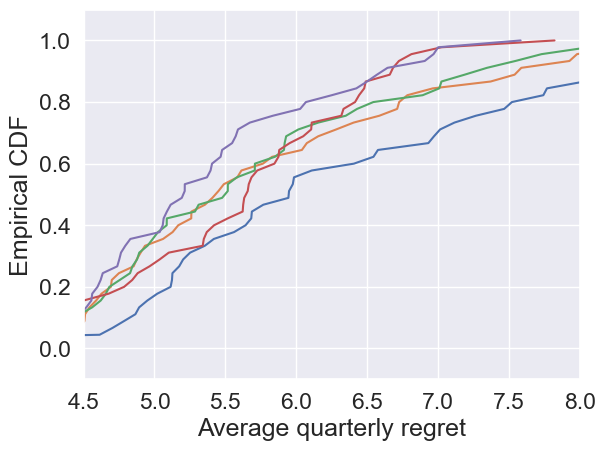}}
\\ \subfigure[Factor data
set.]{\includegraphics[height=0.33\textwidth]{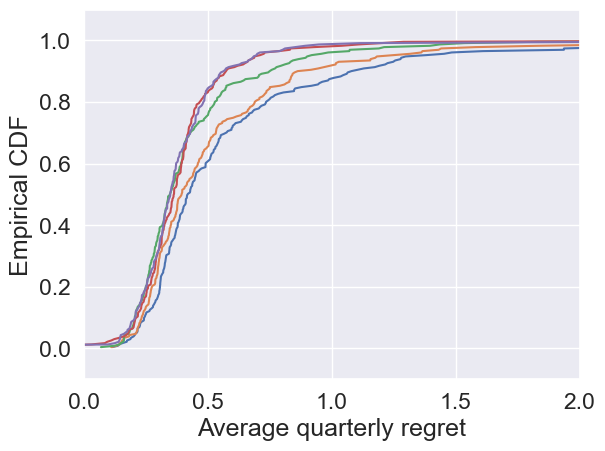}} 
\caption{Cumulative distribution functions of average quarterly regret for five covariance predictors on three data sets.}
\label{f-cdf}
\end{figure}

Figure~\ref{f-cdf} gives a final illustration of these results, by showing the
cumulative distribution functions of the average quarterly regret for the
different covariance predictors.
Clearly, CM-IEWMA has the lowest regret on the industry and stock data set, and
MGARCH does similar on the factor data.

\clearpage 
\section{Portfolio performance} \label{s-portfolio-experiments}
In this section we evaluate the covariance predictors on the portfolios
described in \S\ref{s-pp}. In the minimum variance and mean-variance portfolios,
we use $L_\text{max}=1.6$ (which corresponds to 130:30 long:short),
$w_\text{min} = -0.1$ and $w_\text{max}=0.15$ for the industry and stock return
portfolios, and $w_\text{min} = -0.3$ and $w_\text{max}=0.4$ for the factor
return portfolio. We use target (annualized) volatilities of 5\%, 10\%, and 2\%
for the industry, stock, and factor return portfolios, respectively. 

For the mean-variance portfolio, our estimated returns are EWMAs of the trailing
realized returns. For the industry and stock data we use 250-day half-life EWMAs,
winsorized at the 40th and 60th percentiles (cross-sectionally), and for the factor data a 63-day
half-life EWMA (not winsorized).

\paragraph{Equal weight portfolio.}
Table~\ref{t-eq-w-metrics} shows the metrics for the equal weight portfolio. All
predictors track the volatility targets well. MGARCH attains the highest Sharpe
ratios, although the results are very
close. The drawdowns are also very similar for all predictors, but MGARCH and
CM-IEWMA seem slightly better than the rest.
\begin{table}
\centering
\caption{Metrics for the equal weight portfolio performance for six covariance predictors over the evaluation periods on three data sets.}
\label{t-eq-w-metrics}
\begin{subtable}
\centering
\begin{tabular}{lcccc}
\toprule
{Predictor} & {Return/\%} & {Risk/\%} & {Sharpe} & {Drawdown/\%} \\
\midrule
RW & 2.2 & 5.4 & 0.4 & 16 \\
EWMA & 2.2 & 5.1 & 0.4 & 15 \\
IEWMA & 2.2 & 5.1 & 0.4 & 15 \\
MGARCH & 2.4 & 5.1 & \textbf{0.5} & 14 \\
CM-IEWMA & 2.3 & 5.0\ & \textbf{0.5} & \textbf{13} \\
\hline
PRESCIENT & 4.3 & 4.9 & 0.9 & 8 \\
\bottomrule
\end{tabular}
\captionof*{subtable}{Industry data set.}
\end{subtable}

\vspace{1em}

\begin{subtable}
\centering
\begin{tabular}{lcccc}
\toprule
{Predictor} & {Return/\%} & {Risk/\%} & {Sharpe} & {Drawdown/\%} \\
\midrule
RW & 6.8 & 10.6 & 0.6 & 23 \\
EWMA & 6.4 & 10.0 & 0.6 & 21 \\
IEWMA & 6.7 & 10.1 & 0.7 & 20 \\
MGARCH & 7.2 & 9.4 & \textbf{0.8} & \textbf{15} \\
CM-IEWMA & 6.8 & 9.6 & 0.7 & 17 \\
\hline
PRESCIENT & 12.8 & 9.9 & 1.3 & 10 \\
\bottomrule
\end{tabular}

\captionof*{subtable}{Stock data set.}
\end{subtable}

\vspace{1em}

\begin{subtable}
\centering
\begin{tabular}{lcccc}
\toprule
{Predictor} & {Return/\%} & {Risk/\%} & {Sharpe} & {Drawdown/\%} \\
\midrule
RW & 2.9 & 2.1 & 1.4 & 15 \\
EWMA & 2.9 & 2.0 & 1.4 & 15 \\
IEWMA & 3.0 & 2.0 & 1.5 & 14 \\
MGARCH & 3.2 & 2.0 & \textbf{1.6} & \textbf{12} \\
CM-IEWMA & 2.9 & 2.1 & 1.4 & 15 \\
\hline
PRESCIENT & 3.3 & 2.0 & 1.7 & 12 \\
\bottomrule
\end{tabular}

\captionof*{subtable}{Factor data set.}
\end{subtable}
\end{table}


\paragraph{Minimum variance portfolio.}
Table~\ref{t-min-var-metrics} shows the metrics for the minimum variance
portfolio. For the factor data set, MGARCH does best. On the industry and stock data sets, the three EWMA-based predictors
track the volatility target fairly well, while RW and MGARCH underestimate
volatility. CM-IEWMA and MGARCH both attain a high Sharpe ratio. However, we
note that the high Sharpe ratio for MGARCH, as compared to the other predictors, is
a consequence of the high volatility. Finally, CM-IEWMA seems to consistently
attain a lower drawdown than the other predictors, although the other EWMA-based
approaches also do well.

\begin{table}
\centering
\caption{Metrics for the minimum variance portfolio performance for six covariance predictors over the evaluation periods on three data sets.}
\label{t-min-var-metrics}
\begin{subtable}
\centering
\begin{tabular}{lcccc}
\toprule
{Predictor} & {Return/\%} & {Risk/\%} & {Sharpe} & {Drawdown/\%} \\
\midrule
RW & 3.1 & 5.8 & 0.5 & 23 \\
EWMA & 3.1 & 5.4 & 0.6 & \textbf{19} \\
IEWMA & 3.3 & 5.5 & 0.6 & \textbf{19} \\
MGARCH & 4.3 & 6.1 & \textbf{0.7} & 20 \\
CM-IEWMA & 3.5 & 5.3 & \textbf{0.7} & 20 \\
\hline
PRESCIENT & 3.8 & 5.0 & 0.8 & 13 \\
\bottomrule
\end{tabular}

\captionof*{subtable}{Industry data set.}
\end{subtable}

\vspace{1em}

\begin{subtable}
\centering
\begin{tabular}{lcccc}
\toprule
{Predictor} & {Return/\%} & {Risk/\%} & {Sharpe} & {Drawdown/\%} \\
\midrule
RW & 9.7 & 12.0 & 0.8 & 23 \\
EWMA & 8.9 & 11.1 & 0.8 & 20 \\
IEWMA & 9.7 & 11.3 & \textbf{0.9} & 19 \\
MGARCH & 11.3 & 12.3 & \textbf{0.9} & 18 \\
CM-IEWMA & 9.1 & 11.0 & 0.8 & \textbf{15} \\
\hline
PRESCIENT & 15.6 & 10.0 & 1.6 & 10 \\
\bottomrule
\end{tabular}

\captionof*{subtable}{Stock data set.}
\end{subtable}

\vspace{1em}

\begin{subtable}
\centering
\begin{tabular}{lcccc}
\toprule
{Predictor} & {Return/\%} & {Risk/\%} & {Sharpe} & {Drawdown/\%} \\
\midrule
RW & 1.3 & 2.2 & 0.6 & 20 \\
EWMA & 1.4 & 2.1 & 0.7 & 18 \\
IEWMA & 1.2 & 2.1 & 0.6 & 17 \\
MGARCH & 1.8 & 2.1 & \textbf{0.9} & \textbf{15} \\
CM-IEWMA & 1.2 & 2.1 & 0.5 & 21 \\
\hline
PRESCIENT & 1.0 & 2.0 & 0.5 & 22 \\
\bottomrule
\end{tabular}
\captionof*{subtable}{Factor data set.}
\end{subtable}
\end{table}

\begin{figure}
\centering
\subfigure[Industry
data.]{\includegraphics[height=0.33\textwidth]{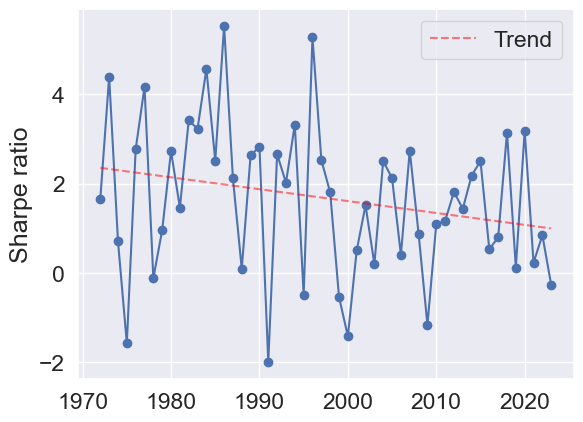}}
\\\subfigure[Stock data
set.]{\includegraphics[height=0.33\textwidth]{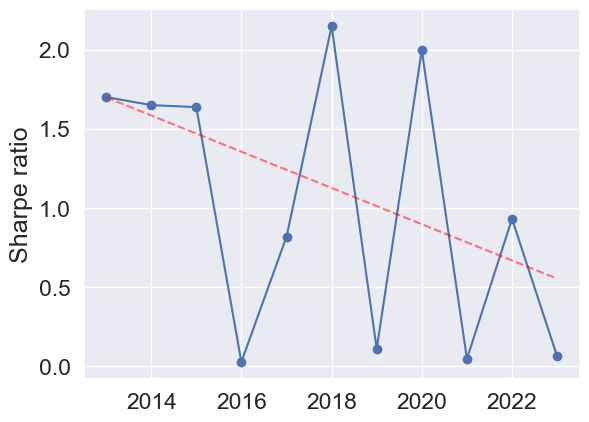}}
\\ \subfigure[Factor data
set.]{\includegraphics[height=0.33\textwidth]{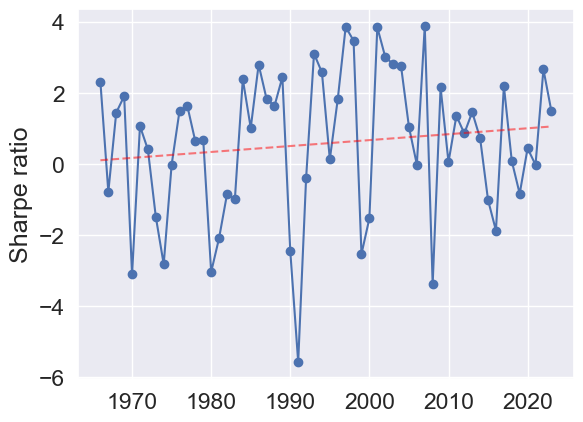}} 
\caption{Yearly annualized Sharpe ratios together with the linear trend 
for minimum variance portfolios on three data sets.}
\label{f-yearly_sr_min_var}
\end{figure}

To illustrate how the minimum variance trading strategy has evolved over time,
we show the yearly annualized Sharpe ratios for the CM-IEWMA predictor in
figure~\ref{f-yearly_sr_min_var}.
We can see that the Sharpe ratio achieved by the 
minimum variance portfolio decreases over time for the industry and stock data sets,
with a small upward trend for the factor data set.

\paragraph{Risk parity portfolio.}
The results for the risk-parity portfolio are shown in
table~\ref{t-risk-par-metrics}. Overall the results are similar for the various
predictors. There is very little that separates the predictors on the industry
data set. On the
stock data, CM-IEWMA and MGARCH attain the highest Sharpe ratios and lowest drawdowns. On the factor data set,
MGARCH has the best overall performance.

\begin{table}
\centering
\caption{Metrics for the risk parity portfolio performance for six covariance predictors over the evaluation periods on three data sets.}
\label{t-risk-par-metrics}
\begin{subtable}
\centering
\begin{tabular}{lcccc}
\toprule
{Predictor} & {Return/\%} & {Risk/\%} & {Sharpe} & {Drawdown/\%} \\
\midrule
RW & 2.4 & 5.4 & \textbf{0.5} & 16 \\
EWMA & 2.4 & 5.1 & \textbf{0.5} & 15 \\
IEWMA & 2.5 & 5.1 & \textbf{0.5} & 14 \\
MGARCH & 2.7 & 5.1 & \textbf{0.5} & 14 \\
CM-IEWMA & 2.5 & 5.0 & \textbf{0.5} & \textbf{13} \\
\hline
PRESCIENT & 4.7 & 4.9 & 1.0 & 8 \\
\bottomrule
\end{tabular}

\captionof*{subtable}{Industry data set.}
\end{subtable}

\vspace{1em}

\begin{subtable}
\centering
\begin{tabular}{lcccc}
\toprule
{Predictor} & {Return/\%} & {Risk/\%} & {Sharpe} & {Drawdown/\%} \\
\midrule
RW & 7.4 & 10.8 & 0.7 & 22 \\
EWMA & 6.8 & 10.1 & 0.7 & 21 \\
IEWMA & 7.2 & 10.2 & 0.7 & 20 \\
MGARCH & 7.9 & 9.7 & \textbf{0.8} & \textbf{15} \\
CM-IEWMA & 7.4 & 9.7 & \textbf{0.8} & 16 \\
\hline
PRESCIENT & 14.3 & 9.9 & 1.5 & 9 \\
\bottomrule
\end{tabular}

\captionof*{subtable}{Stock data set.}
\end{subtable}

\vspace{1em}

\begin{subtable}
\centering
\begin{tabular}{lcccc}
\toprule
{Predictor} & {Return/\%} & {Risk/\%} & {Sharpe} & {Drawdown/\%} \\
\midrule
RW & 1.6 & 2.1 & 0.7 & 19 \\
EWMA & 1.7 & 2.1 & 0.8 & 18 \\
IEWMA & 1.6 & 2.1 & 0.8 & 18 \\
MGARCH & 2.0 & 2.1 & \textbf{1.0} & \textbf{16} \\
CM-IEWMA & 1.5 & 2.1 & 0.7 & 17 \\
\hline
PRESCIENT & 1.4 & 2.0 & 0.7 & 17 \\
\bottomrule
\end{tabular}

\captionof*{subtable}{Factor data set.}
\end{subtable}
\end{table}

\paragraph{Maximum diversification portfolio.}
The maximum diversification portfolio results are illustrated in
table~\ref{t-max-div-metrics}. On the industry and stock data sets, CM-IEWMA and MGARCH do
best in terms of Sharpe ratio, drawdown, and tracking the volatility
target. On the factor data set, MGARCH
does best overall.

\begin{table}
\centering
\caption{Metrics for the maximum diversification portfolio performance for six covariance predictors over the evaluation periods on three data sets.}
\label{t-max-div-metrics}
\begin{subtable}
\centering
\begin{tabular}{lcccc}
\toprule
{Predictor} & {Return/\%} & {Risk/\%} & {Sharpe} & {Drawdown/\%} \\
\midrule
RW & 2.1 & 5.5 & 0.4 & 16 \\
EWMA & 2.1 & 5.1 & 0.4 & 16 \\
IEWMA & 2.2 & 5.2 & 0.4 & 14 \\
MGARCH & 2.5 & 5.1 & \textbf{0.5} & \textbf{12} \\
CM-IEWMA & 2.3 & 5.0 & \textbf{0.5} & \textbf{12} \\
\hline
PRESCIENT & 3.8 & 5.0 & 0.8 & 10 \\
\bottomrule
\end{tabular}

\captionof*{subtable}{Industry data set.}
\end{subtable}

\vspace{1em}

\begin{subtable}
\centering
\begin{tabular}{lcccc}
\toprule
{Predictor} & {Return/\%} & {Risk/\%} & {Sharpe} & {Drawdown/\%} \\
\midrule
RW & 8.4 & 11.2 & 0.8 & 22 \\
EWMA & 7.9 & 10.4 & 0.8 & 21 \\
IEWMA & 8.2 & 10.4 & 0.8 & 20 \\
MGARCH & 10.0 & 9.8 & \textbf{1.0} & \textbf{15} \\
CM-IEWMA & 8.8 & 10.0 & 0.9 & 16 \\
\hline
PRESCIENT & 13.5 & 9.9 & 1.4 & 11 \\
\bottomrule
\end{tabular}

\captionof*{subtable}{Stock data set.}
\end{subtable}

\vspace{1em}

\begin{subtable}
\centering
\begin{tabular}{lcccc}
\toprule
{Predictor} & {Return/\%} & {Risk/\%} & {Sharpe} & {Drawdown/\%} \\
\midrule
RW & 1.4 & 2.2 & 0.7 & 19 \\
EWMA & 1.5 & 2.1 & 0.7 & 19 \\
IEWMA & 1.4 & 2.1 & 0.7 & 19 \\
MGARCH & 2.0 & 2.1 & \textbf{1.0} & \textbf{16} \\
CM-IEWMA & 1.4 & 2.1 & 0.7 & 18 \\
\hline
PRESCIENT & 1.3 & 2.0 & 0.7 & 18 \\
\bottomrule
\end{tabular}

\captionof*{subtable}{Factor data set.}
\end{subtable}
\end{table}

\paragraph{Mean variance portfolio.}
The results for the mean-variance portfolio are given in
table~\ref{t-mean-var-metrics}. On the industry data set all predictors
underestimate volatility. The results are similar across predictors, with
CM-IEWMA and MGARCH performing slightly better than the rest in terms of Sharpe ratio and
drawdown. On the stock data set, CM-IEWMA
seems to do best overall. On the factor data set, the results are almost
identical between predictors.
\begin{table}
\centering
\caption[width=0.1\textwidth]{Metrics for the mean variance portfolio performance for six covariance predictors over the evaluation periods on three data sets.}
\label{t-mean-var-metrics}
\begin{subtable}
\centering
\begin{tabular}{lcccc}
\toprule
{Predictor} & {Return/\%} & {Risk/\%} & {Sharpe} & {Drawdown/\%} \\
\midrule
RW & 5.6 & 6.2 & 0.9 & 16 \\
EWMA & 5.6 & 5.8 & 1.0 & 15 \\
IEWMA & 5.9 & 5.7 & 1.0 & 14 \\
MGARCH & 6.7 & 6.4 & 1.0 & 14 \\
CM-IEWMA & 6.1 & 5.6 & \textbf{1.1} & \textbf{13} \\
\hline
PRESCIENT & 4.6 & 5.0 & 0.9 & 10 \\
\bottomrule
\end{tabular}

\captionof*{subtable}{Industry data set.}
\end{subtable}

\vspace{1em}

\begin{subtable}
\centering
\begin{tabular}{lcccc}
  \toprule
  {Predictor} & {Return/\%} & {Risk/\%} & {Sharpe} & {Drawdown/\%} \\
  \midrule
  RW & 6.1 & 11.9 & 0.5 & 26 \\
  EWMA & 5.9 & 11.0 & 0.5 & 20 \\
  IEWMA & 7.9 & 11.1 & \textbf{0.7} & 15 \\
  MGARCH & 8.3 & 11.9 & \textbf{0.7} & 18 \\
  CM-IEWMA & 7.3 & 10.9 & \textbf{0.7} & \textbf{13} \\
  \hline
  PRESCIENT & 14.3 & 9.9 & 1.4 & 9 \\
  \bottomrule
\end{tabular}

\captionof*{subtable}{Stock data set.}
\end{subtable}

\vspace{1em}

\begin{subtable}
\centering
\begin{tabular}{lcccc}
\toprule
{Predictor} & {Return/\%} & {Risk/\%} & {Sharpe} & {Drawdown/\%} \\
\midrule
RW & 7.5 & 2.2 & 3.3 & 4 \\
EWMA & 7.2 & 2.1 & \textbf{3.4} & 4 \\
IEWMA & 7.1 & 2.1 & 3.3 & 4 \\
MGARCH & 7.3 & 2.2 & 3.3 & \textbf{3} \\
CM-IEWMA & 6.9 & 2.2 & 3.2 & 4 \\
\hline
PRESCIENT & 6.5 & 1.9 & 3.3 & 4 \\
\bottomrule
\end{tabular}
\captionof*{subtable}{Factor data set.}
\end{subtable}
\end{table}

\begin{figure}
\centering
\subfigure[Industry
data.]{\includegraphics[height=0.33\textwidth]{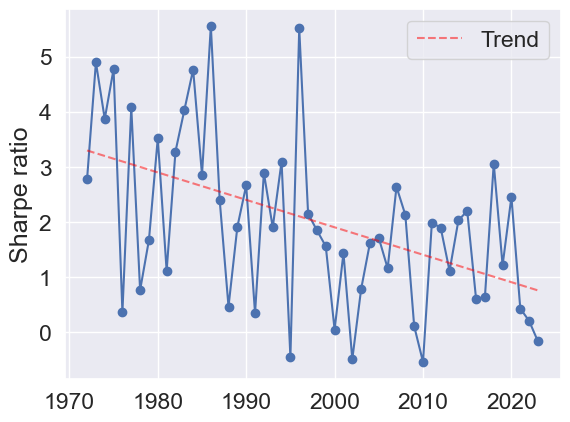}}
\\\subfigure[Stock data
set.]{\includegraphics[height=0.33\textwidth]{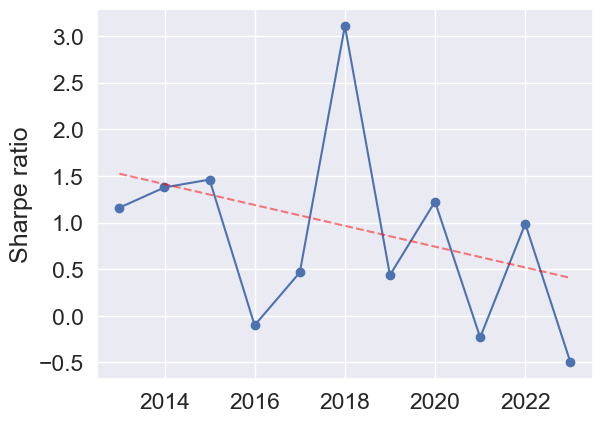}}
\\ \subfigure[Factor data
set.]{\includegraphics[height=0.33\textwidth]{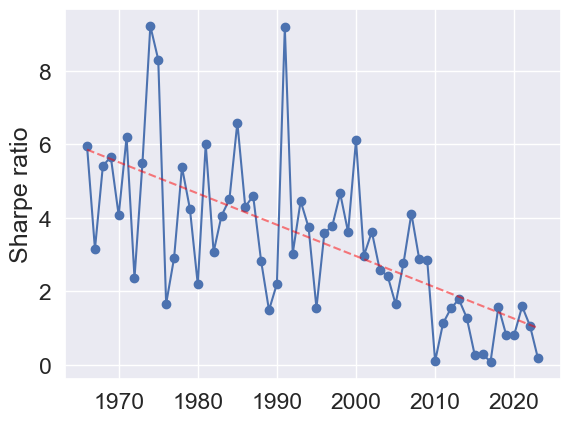}} 
\caption{Yearly annualized Sharpe ratios together with the linear trend for mean variance portfolios on three data sets.}
\label{f-yearly_sr_mean_var}
\end{figure}

Since we use simple EWMA return predictors, we can expect the mean variance
portfolio performance to vary over time. Intuitively it should be better on
historical data than more recent data. To illustrate this,
figure~\ref{f-yearly_sr_mean_var} shows the yearly annualized Sharpe ratios for
the three portfolios.
There is a clear downward trend in the Sharpe ratios for the industry and factor
data sets, illustrating the difficulty of predicting returns in recent years.
This can be compared to the minimum variance portfolios
(figure~\ref{f-yearly_sr_min_var}) that have a more stable performance over
time, and notably do not depend on a mean estimate.

\clearpage 
\section{Summary}
In terms of log-likelihood and regret, CM-IEWMA performs best, followed by
MGARCH, which performs better than the simpler covariance predictors. In downstream portfolio optimization experiments, CM-IEWMA and MGARCH again
perform better than the other predictors, although in many cases not by much. In these experiments there is more variation in the results,
partly explained by the difference between our prediction (of a covariance
matrix) and our metrics (such as return, risk, drawdown). Even the simplest
covariance predictors do a reasonable job of predicting the portfolio risk.

\chapter{Realized covariance} \label{c-realized-cov}
We have so far focused on predicting the covariance matrix of asset 
returns using historical return data, \ie, we predict $\sigmahat_t$ from
$r_1, \ldots, r_{t-1}$.
In this chapter we consider the use of additional data, specifically, 
intraperiod returns.
As an example, suppose the period is (trading) days.  The methods
described in previous chapters predict the covariance of the daily return
from previous daily returns.
In so-called realized covariance, we predict the daily return covariance
using intraday returns.
Instead of single period returns $r_t$, we have multiple returns associated
with period $t$.
It is not surprising that using multiple realized returns for each period,
instead of just one, can improve our covariance estimates. 

Recent literature has shown that realized volatility and correlation
measurements (based on high-frequency intraperiod data) can
improve performance over traditional predictors that rely on a single
realization per period. \citet{hansen2012realized} extend the univariate GARCH model to
the joint modeling of returns and realized measures of volatility, and show
empirically that this improves performance over the standard GARCH model. In
\citep{bauwens2012dynamic} a multivariate realized GARCH model is proposed. More
recently, \citep{bollerslev2020multivariate} propose a realized semicovariance
GARCH model to allow for nuanced responses to positive and negative return
shocks.

In this chapter we show that the dynamically weighted prediction combiner of
\S\ref{s-dynamic-combiner} readily handles multiple realized returns per period.
For simplicity we will assume each period has the same number of intraperiod returns,
equally spaced in time.
We redefine the return vector to be a return matrix
$r_t\in \reals^{n\times m}$ with columns that are the
$m$ intraperiod return vectors, for times $t=1,\ldots,T$. 
The realized covariance at time $t$ is defined as
\[
C_t = r_tr_t^T,
\]
the same formula for the realized return when $r_t$ is a single (vector) return.
The realized covariance matrix $C_t$ has rank $m$ when the $m$ return vectors
are linearly independent and $m\leq n$; this can be compared to the realized covariance
when we do not have intraperiod returns, which is rank one.

\section{Combined multiple realized EWMAs}
The dynamically weighted prediction combiner of \S\ref{s-dynamic-combiner}
readily handles multiple realized covariance predictors. 
\paragraph{Realized EWMA.}
We define the realized EWMA (REWMA) predictor as
\[ 
\sigmahat_t = \alpha_{t} \sum_{\tau=1}^{t-1}\beta^{t-1-\tau} C_\tau,
\quad t=2,3, \ldots,
\]
where $C_\tau$ is the realized covariance at time $\tau$, $\beta
\in(0,1)$ is the forgetting factor, and $\alpha_t$ is the normalizing constant;
see \S\ref{s-ewma} for details.
This is the same formula as the usual EWMA covariance, with one return per period,
given in \eqref{e-ewma}, with $r_t$ extended to be a matrix of multiple returns.

\paragraph{Combined multiple realized EWMAs.}
The combined multiple realized EWMA (CM-REWMA) predictor starts with a set of $K$ REWMA
predictors $\sigmahat_t^{(k)}$ with half-lives $H^{(k)}$, $k=1,\ldots,K$, and
combines them using the dynamically weighted prediction combiner of \S
\ref{s-dynamic-combiner}.

\section{Data and experimental setup}
\paragraph{Data set.}
We consider a universe of $n=39$ assets with five-minute intraday returns corresponding
to $m=77$. The assets were taken as a subset of those used
by~\citet{pelger2020understanding}, and are available at~\citep{pelger-website}.
The data set spans January 2nd 2004 to December 30th 2016, for a total of 252021
data points over 3273 trading days. We list the assets in table~\ref{tab:realized-tickers}.
\begin{table}
\small
\centering
\begin{tabular}{cc}
\toprule
Ticker & Company Name \\
\midrule
JPM & JPMorgan Chase  \\
GS & Goldman Sachs  \\
KO & The Coca-Cola Company  \\
IBM & International Business Machines Corporation  \\
CAT & Caterpillar Inc.  \\
CVX & Chevron Corporation  \\
XOM & Exxon Mobil Corporation  \\
GE & General Electric Company  \\
MRK & Merck \& Co., Inc.  \\
VZ & Verizon Communications Inc.  \\
PFE & Pfizer Inc.  \\
WMT & Walmart Inc.  \\
C & Citigroup Inc.  \\
HD & The Home Depot, Inc.  \\
BA & The Boeing Company  \\
MMM & 3M Company  \\
MCD & McDonald's Corporation  \\
NKE & NIKE, Inc.  \\
JNJ & Johnson \& Johnson  \\
INTC & Intel Corporation  \\
MSFT & Microsoft Corporation  \\
AAPL & Apple Inc.  \\
AMZN & Amazon.com Inc.  \\
CSCO & Cisco Systems, Inc.  \\
PG & Procter \& Gamble Co.  \\
ABT & Abbott Laboratories  \\
VLO & Valero Energy Corporation  \\
HON & Honeywell International Inc.  \\
LMT & Lockheed Martin Corporation  \\
TXN & Texas Instruments Inc.  \\
COST & Costco Wholesale Corporation  \\
PEP & PepsiCo, Inc.  \\
UNP & Union Pacific Corporation  \\
WFC & Wells Fargo \& Co.  \\
CVS & CVS Health Corporation  \\
ORCL & Oracle Corporation  \\
XRX & Xerox Corporation  \\
TMO & Thermo Fisher Scientific Inc.  \\
NSC & Norfolk Southern Corporation  \\
\bottomrule
\end{tabular}
\caption{Assets used in realized covariance study.}
\label{tab:realized-tickers}
\end{table}

\paragraph{Four covariance predictors.}
We evaluate four covariance predictors, described below. 
\BIT
\item The CM-IEWMA predictor used for the stock data from
\S\ref{s-experiment-predictors}. This predictor
only uses daily returns, and is not a realized covariance predictor.
\item An REWMA predictor with a half-life of $H=10$ days, denoted REWMA-10.
\item A CM-REMWA predictor with five components with half-lives of $1, 5, 10,
21$, and $63$ days, respectively.
\item Prescient predictor, \ie, the empirical covariance for the quarter the day
is in. As with the CM-IEWMA predictor, this predictor uses daily return data.
It is of course not implementable, and meant only to show a bound
on performance with which to compare our implementable predictors.
\EIT

\section{Empirical results}
\begin{figure}
\centering
\includegraphics[width=0.8\textwidth]{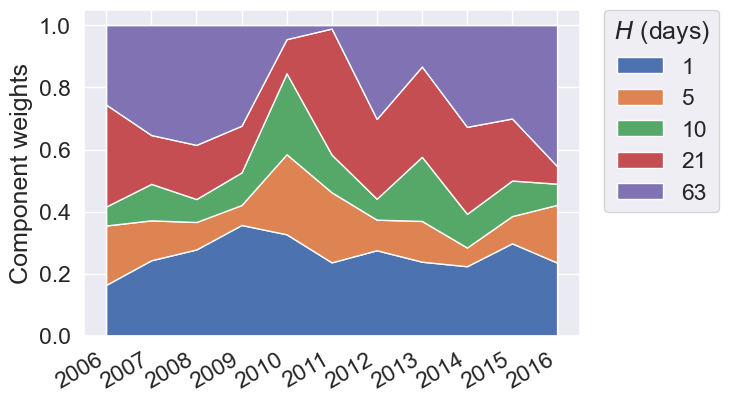}
\caption{CM-REWMA component weights, averaged annually.}
\label{fig:realized-weights}
\end{figure}

\paragraph{CM-REWMA component weights.}
Figure~\ref{fig:realized-weights} shows the component weights for the CM-REWMA
predictor, averaged annually.
The weights are fairly stable over time but a weight shift toward
faster changing EWMAs is seen in 2008, during the financial crisis.

\paragraph{MSE.}
The average, standard deviation, and maximum MSEs, computed over distinct
quarters for the four covariance predictors, are given in table~\ref{tab:realized-mse}.
\begin{table}[ht]
\centering
\begin{tabular}{lccc}
\toprule
Predictor & Average/$10^{-4}$ & Std. Dev./$10^{-3}$ & Max/$10^{-3}$ \\
\midrule
CM-IEWMA & $3.1$ & $1.2$ & $7.6$ \\
REWMA & $\mathbf{3.0}$ & $\mathbf{1.1}$ & $7.3$ \\
CM-REWMA & $\mathbf{3.0}$ & $\mathbf{1.1}$ &
$\mathbf{7.2}$ \\
\hline
PRESCIENT & $3.0$ & $1.1$ & $7.1$ \\
\bottomrule
\end{tabular}
\caption{Metrics on the MSE, computed over distinct quarters.}
\label{tab:realized-mse}
\end{table}
The REWMA and CM-REWMA do slightly better than the CM-IEWMA predictor, but
overall there is not a big difference between the predictors. 

\begin{figure}
\centering
\includegraphics[width=0.8\textwidth]{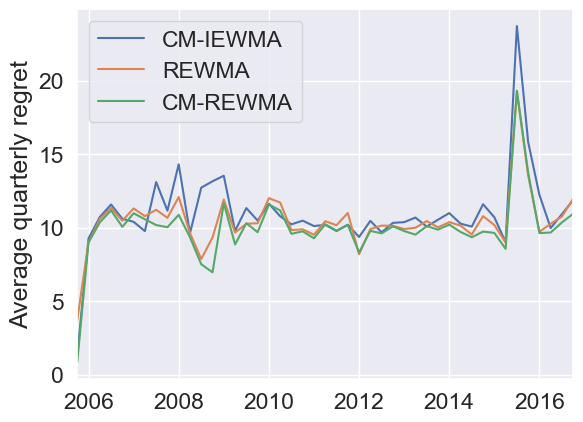}
\caption{Average regret over distinct quarters for three covariance predictors.}
\label{fig:realized-regret}
\end{figure}

\paragraph{Regret.}
Figure~\ref{fig:realized-regret} shows the average regret over distinct quarters
for the CM-IEWMA, REWMA, and CM-REWMA predictors.
The CM-REWMA predictor has the lowest regret in almost all quarters. It has
lower regret than the REWMA predictor in 41 out of the 50 quarters, and lower
regret than the CM-IEWMA predictor in 39 out of the 50 quarters. 

\begin{figure}
\centering
\includegraphics[width=0.8\textwidth]{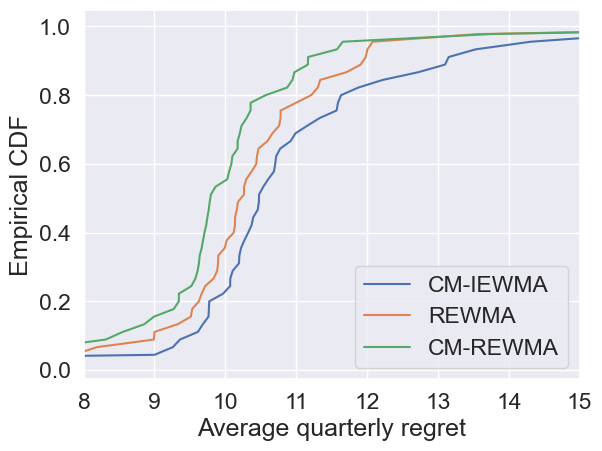}
\caption{Cumulative distribution functions of the average quarterly regret for
three covariance predictors.}
\label{f-realized-cdf}
\end{figure}

Finally, figure~\ref{f-realized-cdf} shows the
cumulative distribution functions of the average quarterly regret for the
different covariance predictors.
CM-REWMA has lower regret than both the CM-IEWMA and REWMA predictors, while
REWMA has lower regret than CM-IEWMA.

\paragraph{Portfolio performance.}
Table~\ref{t-realized-portfolio-metrics} shows the portfolio metrics for five
different portfolio construction methods.
\begin{table}
\centering
\begin{tabular}{lcccc}
\toprule
{Predictor} & {Return/\%} & {Risk/\%} & {Sharpe} & {Drawdown/\%} \\
\midrule
\multicolumn{5}{l}{\textbf{Equal weight}} \\
CM-IEWMA & 3.2 & 9.9 & 0.3 & 18 \\
REWMA & 3.7 & 10.3 & \textbf{0.4} & \textbf{16} \\
CM-REWMA & 4.4 & 10.6 & \textbf{0.4} & \textbf{16} \\
\hline
PRESCIENT & 6.7 & 9.9 & 0.7 & 13 \\
\midrule
\multicolumn{5}{l}{\textbf{Minimum variance}} \\
CM-IEWMA & 10.7 & 11.0 & 1.0 & 25 \\
REWMA & 10.7 & 10.5 & 1.0 & \textbf{21} \\
CM-REWMA & 12.0 & 10.7 & \textbf{1.1} & \textbf{21} \\
\hline
PRESCIENT & 11.7 & 10.0 & 1.2 & 12 \\
\midrule
\multicolumn{5}{l}{\textbf{Risk parity}} \\
CM-IEWMA & 4.1 & 10.0 & 0.4 & 18 \\
REWMA & 4.7 & 10.3 & \textbf{0.5} & \textbf{17} \\
CM-REWMA & 5.5 & 10.6 & \textbf{0.5} & \textbf{17} \\
\hline
PRESCIENT & 8.0 & 9.9 & 0.8 & 12 \\
\midrule
\multicolumn{5}{l}{\textbf{Maximum diversification}} \\
CM-IEWMA & 3.6 & 10.2 & 0.4 & 25 \\
REWMA & 4.3 & 10.5 & 0.4 & 21 \\
CM-REWMA & 5.1 & 10.8 & \textbf{0.5} & \textbf{19} \\
\hline
PRESCIENT & 7.8 & 9.9 & 0.8 & 16 \\
\midrule
\multicolumn{5}{l}{\textbf{Mean variance}} \\
CM-IEWMA & 8.6 & 10.5 & 0.8 & 22 \\
REWMA & 8.5 & 10.3 & 0.8 & \textbf{16} \\
CM-REWMA & 9.3 & 10.5 & \textbf{0.9} & 19 \\
\hline
PRESCIENT & 10.9 & 9.8 & 1.1 & 21 \\
\bottomrule
\end{tabular}
\caption{Metrics for five different portfolio construction methods, using four covariance predictors.}
\label{t-realized-portfolio-metrics}
\end{table}
CM-REWMA does better than, or as well as, REWMA on almost all metrics, and
better than CM-IEWMA for all portfolios. However, the difference on
portfolio tasks is not large.

\paragraph{Summary.}
The results above show that using realized covariance, \ie, intraperiod returns
instead of just one return per period, gives covariance estimates that are a bit 
better than those obtained using only one return per period.

\chapter{Large universes} \label{c-large-universes}
In a practical setting we often encounter a larger number of assets than
considered in the previous chapters, which has led to extensive research in
high-dimensional covariance estimation.
One challenge in large dimensions is ensuring positive definiteness of the
covariance matrix, in particular with model-based approaches such as
MGARCH~\citep{garch_survey}. Several techniques have been proposed for
estimating MGARCH models in large dimensions; see, \eg, \citep{engle2019large,
de2021factor, de2022large}. Others have focused on estimating
realized covariance matrices in high dimensions; see, \eg, \citep{oh2016high,
vassallo2021dcc, hautsch2015high, debrito2018forecasting, fan2016incorporating,
ait2017using}. For a detailed review of recent developments in high-dimensional
covariance estimation, we recommend~\citep[\S 6]{bauwens2023modeling}.

The methods described in previous chapters can be adapted to handle large universes of assets,
say $n$ larger than $100$ or so.  In this chapter we describe two closely 
related methods for improving the performance with large $n$.
Both methods end up modeling $\hat \Sigma_t$ as a low rank plus diagonal matrix,
in so-called factor form. Before describing these methods, we mention that evaluating log-likelihood
regret is complicated with large $n$.  For the empirical covariance to be nonsingular
(which is needed to evaluate the regret),
we need at least $n$ periods; for daily returns with $n=1000$, this amounts
to four years. Even if we have $n$ periods of data, we would only be able to
evaluate the regret a few times. For example, with $n=1000$ (four years) we need
at least 40 years of data to compute the average regret over 10 distinct
periods. 
The log-likelihood, however, can still be evaluated over fewer than
$n$ periods. 

\section{Traditional factor model} 
In practice most return covariance matrices for large universes are constructed
from factors, with the model
\[
r_t = F_t f_t + z_t, \quad t=1,2, \ldots,
\]
where $F_t \in \reals^{n \times k}$ is the factor exposure matrix, $f_t \in
\reals^k$ is the factor return vector, $z_t \in \reals^n$ is the idiosyncratic
return, and $k$ is the number of factors, typically much smaller than $n$. 
The factor returns are constructed or found by several methods, 
such as principal component analysis (PCA), or by hand; see, 
\eg, \citep{bai2008large, bai2003inferential,
lettau2020estimating, lettau2020factors, pelger2022factor,
pelger2022interpretable, fama1993common, fama1992cross}. 
Thus we assume that the factor returns are known.  Given the factor
returns, the rows of the factor exposure matrix are typically found by least
squares regression over a rolling or exponentially weighted
window~\citep{cochrane2009asset}. The idiosyncratic returns $z_t$ are then found
as the residuals in this least squares fit.
The factor returns $f_t$ are modeled as $\mathcal N(0,\Sigma_t^\mathrm{f})$,
and the idiosyncratic returns $z_t$ are modeled as $\mathcal N(0,E_t)$,
where $E_t$ is diagonal.  It is also assumed that the factor returns and
idiosyncratic returns are independent across time and of each other.

We end up with a covariance matrix in factor form, \ie, rank $k$ plus diagonal,
\BEQ\label{e-factor-model-cov}
\Sigma_t = F_t\Sigma^{\text{f}}_tF_t^T + {E}_t.
\EEQ
We can easily use the methods described above with a factor model.
Simply predict the factor return covariance $\hat \Sigma^\text{f}_t$
(using the factor returns $f_t$)
and the idiosyncratic variances $\hat E_t$ (using the entries of $z_t$),
using the methods described in this monograph, and then form the covariance estimate 
\[
\sigmahat_t = F_t\hat{\Sigma}^{\text{f}}_tF_t^T + \hat{E}_t.
\]

The factor model \eqref{e-factor-model-cov} can be written in 
a simpler form as 
\BEQ\label{e-factor-model}
\Sigma_t = \tilde F_t \tilde F_t^T + E_t,
\EEQ
with $\tilde F_t = F_t(\Sigma_t^\text{f})^{1/2}$.
This form does not include a factor covariance $\Sigma^\text{f}$, 
or equivalently, assumes $\Sigma^\text{f}_t =I$, \ie, the factors 
are independent with standard deviation one.
(The associated factors are called whitened factors.)
We will use the factor model form \eqref{e-factor-model} in the sequel.

The factor model \eqref{e-factor-model} has parameters $\tilde F_t$ and $E_t$, which
all together include $nk+n$ scalar parameters. (Some of these are
redundant; for example we can insist without loss of generality 
that $F$ is lower triangular.)
The factor model contains substantially fewer scalar parameters 
than a generic $n \times n$ covariance matrix, which contains $n(n+1)/2$ 
scalar parameters.

The smaller number of parameters is not
the only reason for using a factor model.  Another is that it often gives
better covariance estimates.  We can think of the low rank plus diagonal
structure as regularization, which can improve out-of-sample
performance.  In addition, the low rank plus diagonal structure 
can be exploited in portfolio construction, bringing the computational
complexity down from $O(n^3)$ to $O(nk^2)$ operations \citep{boyd2004convex}.
This makes portfolio optimization with $n=1000$ assets and $k=50$ factors
extremely fast, and makes possible optimization of portfolios with
much larger values of $n$.





\section{Fitting a factor model to a covariance matrix}\label{s-factor-fitting}

In this section we consider the problem of fitting a given covariance
matrix $\Sigma$ by one in factor form, $\hat \Sigma = FF^T+ E$, 
where $F \in \reals^{n \times k}$.
This corresponds to the model $r=Ff+z$, with (factor return) $f \sim \mathcal N(0,I)$,
and (idiosyncratic return) $r \sim \mathcal (0,E)$, with $E$ diagonal.
We let $\theta = (F,E)$ denote the parameters of our factor form model.

We seek $F\in \reals^{n \times k}$ and diagonal $E\in \reals^{n \times n}$
(with positive diagonal entries)
that minimize the Kullback-Liebler (KL) divergence between $\mathcal N(0,\Sigma)$
and $\mathcal N(0,\hat \Sigma)$,
\BEQ\label{e-KL}
\mathcal K(\Sigma,\hat \Sigma) = \frac{1}{2}
\left(
\log \frac{\det \hat \Sigma}{\det \Sigma} - n + \Tr \hat\Sigma^{-1} \Sigma
\right).
\EEQ
The KL divergence can be expressed in terms of the average 
log-likelihood of $\mathcal N(0,\hat\Sigma)$ under
$\mathcal N(0,\Sigma)$ as
\BEQ\label{e-em-ll}
\Expect_{r \sim \mathcal{N}(0,\Sigma)}\ell_{\sigmahat}(r) = -\mathcal K(\Sigma,\hat \Sigma) - (1/2)(n\log 2\pi +n + \log\det \Sigma),
\EEQ
where $\ell_{\sigmahat}(r)$ is the log-likelihood of $r$
under $\mathcal N(0,\hat \Sigma)$. Hence minimizing the 
KL-divergence~\eqref{e-KL} is
equivalent to maximizing the expected log-likelihood~\eqref{e-em-ll} 
of $r$ under the model $\mathcal N(0,\hat \Sigma)$.
 
\paragraph{Solution via EM.}
We can use the expectation-maximization (EM) algorithm to approximately 
minimize $\mathcal K(\Sigma,FF^T+E)$ over $F$ and 
$E$~\citep{dempster1977maximum,rubin1982algorithms}.
Usually EM is used to fit a factor model to data, \ie, samples; here we 
use it to fit a given Gaussian distribution $\mathcal N(0,\Sigma)$.
The method described below was suggested and derived by Emmanuel Cand\a`es.
We are not aware of its appearance in prior literature. A forthcoming paper on this method will include more detail and applications.

\def\independenT#1#2{\mathrel{\rlap{$#1#2$}\mkern2mu{#1#2}}}
\newcommand\independent{\protect\mathpalette{\protect\independenT}{\perp}}


The EM algorithm is an iterative method for maximizing \eqref{e-em-ll}.
Each iteration consists of two steps: the expectation or E-step,
and the maximization or M-step.
We use the conventional symbols used to describe EM, and use
subscript $j=1,2, \ldots$ to denote iteration number.
(A good method for initializing the EM algorithm is provided below.)

\paragraph{E-step.}
In the E-step, we find the expected log-likelihood under the 
current estimate of the parameters $\theta_{j} = (F_j,E_j)$, over the
true distribution of $r$:
\BEQ \label{e-Q}
Q(\theta \, || \, \theta_j) = \Expect_{r \sim \mathcal{N}(0,\Sigma)} 
\Expect_{p_{\theta_j}(f \mid r)} \ell_\theta(r,f)  
\EEQ
where $p_{\theta_j}(f \mid r)$ is the density of the conditional distribution
of $f$ under the parameter estimates at iteration $j$, 
and $\ell_{\theta}(r,f)$ is the log likelihood of the joint distribution with variable
$\theta=(F,E)$.

With our factor model the complete log-likelihood of $(r,f)$ is
\BEAS
\ell_\theta(r,f) &=& -\frac{1}{2} \left((r - F f)^T E^{-1} (r - F f) + f^T f + 
\log \det E\right) \\ &&+ \frac{1}{(2\pi)^{n/2}} + \frac{1}{(2\pi)^{k/2}} - k/2.
\EEAS
The conditional distribution of $f\mid r$ under $\theta_j$ is~\citep{bishop2006pattern}
\[
f \mid r ~\sim~ \mathcal{N}(B_j r, G_j),
\]
where
\BEQ\label{e-BG}
B_j = G_j F_j^T E_j^{-1}, \quad G_j^{-1} = F_j^T E_j^{-1} F_j + I. 
\EEQ
Hence, \eqref{e-Q} becomes, up to an additive constant,
\BEQ\label{e-Q-closed-form}
-\frac{1}{2}\Tr(E^{-1}\, (C_{rr} - 2C_{rf} F^T + F C_{ss}
F^T)) - \frac{1}{2}\log\det E, 
\EEQ
where
\BEQ\label{e-Crr}
C_{rr}= \Sigma, \quad C_{rf} = \Sigma B_j^T, \quad C_{ff} = B_j \Sigma B_j^T + G_j.
\EEQ

\paragraph{M-step.}
In the M-step \eqref{e-Q} is maximized with respect to $\theta$ to obtain the
updated parameters:
\[
\theta_{j+1} = \argmax_\theta \,\, Q(\theta \, || \, \theta_j).  
\]
The maximizer of~\eqref{e-Q-closed-form} is~\citep{rubin1982algorithms}
\BEAS
{F}_{j+1} &=& C_{rf} C_{ff}^{-1},\\ 
{E}_{j+1} &=& \diag(\diag(C_{rr} - 2C_{rf} F_{j+1}^T + {F}_{j+1} C_{ff} F_{j+1}^T)),
\EEAS
where the inner $\diag$ extracts the diagonal of its (matrix) argument, and the outer
$\diag$ creates a diagonal matrix from its (vector) argument.

\paragraph{EM iteration.}
The EM iteration has the form
\BEAS
{F}_{j+1} &=& C_{rf} C_{ff}^{-1},\\ 
{E}_{j+1} &=& \diag(\diag(C_{rr} - 2C_{rf} F_{j+1}^T + {F}_{j+1} C_{ff} F_{j+1}^T)),
\EEAS
where $C_{rr}$, $C_{rf}$, and $C_{ff}$ come from \eqref{e-BG} and
\eqref{e-Crr}.

\paragraph{Initialization.}
To initialize the EM algorithm we use the following method, based on low rank
approximation via eigendecomposition.
We work with the correlation matrix of $\Sigma$, denoted 
\[
R = \diag(\sigma)^{-1} \Sigma \diag(\sigma)^{-1},
\]
where $\sigma= \diag(\Sigma)^{1/2}$ (entrywise).
First we express $R$ in its eigendecomposition $R = \sum_{i=1}^n
\lambda_i q_iq_i^T$, with $\lambda_1 \geq \lambda_2 \geq \cdots \geq \lambda_n$.
We then form the rank $k$ approximation
\[
\widehat R = \sum_{i=1}^k \lambda_i q_iq_i^T.
\]
We only need to compute the $k$ dominant eigenvectors and eigenvalues,
which can be done efficiently using for example the Lanczos
algorithm~\citep{golub2013matrix}. Let 
\[
\widehat E = \diag \left( \diag(R-\widehat R) \right),
\]
which can be shown to have positive diagonal entries.
Our low-rank plus diagonal approximation of $R$ is then 
$\widehat R+\widehat E$.
It is also a correlation matrix, \ie, has diagonal entries one.
Our final factor approximation of $\Sigma$ is given by
\[
\diag(\sigma) (\widehat R+\widehat E) \diag(\sigma) = FF^T+E,
\]
with
\[
F = \diag(\sigma) [\sqrt{\lambda_1} q_1 \cdots \sqrt{\lambda_k} q_k], \qquad
E = \diag(e \circ \sigma^2),
\]
where $\circ$ denotes the elementwise (Hadamard) product, and $\sigma^2$
means elementwise.
 
This initialization alone can serve as a basic method to
fit a factor model to a given covariance matrix.  We will see below
that in terms of portfolio optimization, it serves just as well
as a factor model fit using the EM method.

\section{Data and experimental setup}
\paragraph{Data set.}
We gather the 500 largest NASDAQ stocks (by market capitalization) at the beginning of 2000
from the WRDS portal~\citep{WRDS}, compute the daily returns of these stocks
from January 3rd 2000 to December 30th 2022, and remove any stocks with missing
return values during this period. This gives us 238 stocks over 5787 (trading) days.
We acknowledge that we induce a survivor
bias, but the purpose of this empirical study is solely to demonstrate the benefit
of regularization in large universes, and not to backtest a trading strategy.

\paragraph{Traditional factor model.}
We create a factor model using PCA as follows. Every year, the $k$
principal components of largest explanatory power are computed, using the past two years of
returns. These define the columns of the factor exposure matrix $F_t$ for the
following year, and the factor returns $f_t$ are the projections of the returns
onto these principal components. The idiosyncratic returns $z_t$ are the
residuals.  We leverage the CM-IEWMA predictor to compute
the factor covariance, using three IEWMA components with half-lives (in days)
$H^{\text{vol}}/H^{\text{cor}}$ of $\lceil k/2 \rceil/k, k/3k$, and $3k/6k$, where $k$
denotes the number of factors. To estimate the idiosyncratic variances a 21-day
EWMA is used. We evaluate the factor models on the average log-likelihood over the evaluation period.

\paragraph{Fitting a factor model to the covariance matrix.}
We use a CM-IEWMA covariance predictor with four IEWMA components with
half-lives $63/125, 125/250, 250/500$, and  $500/1000$ days, respectively, given as
$H^{\text{vol}}/H^{\text{cor}}$. Given the CM-IEWMA estimate $\sigmahat_t$ at
time $t$, we approximate it using a factor model as described in
\S\ref{s-factor-fitting}.

To evaluate the factor models, we look at the average log-likelihood over the evaluation period and several performance metrics for a minimum variance
portfolio with $L_\text{max}=1.6$,
$w_\text{min} = -0.1$, and $w_\text{max}=0.15$, diluted to a target risk of 10\%.

\section{Empirical results}
\begin{figure}
\centering
\includegraphics[width=0.8\textwidth]{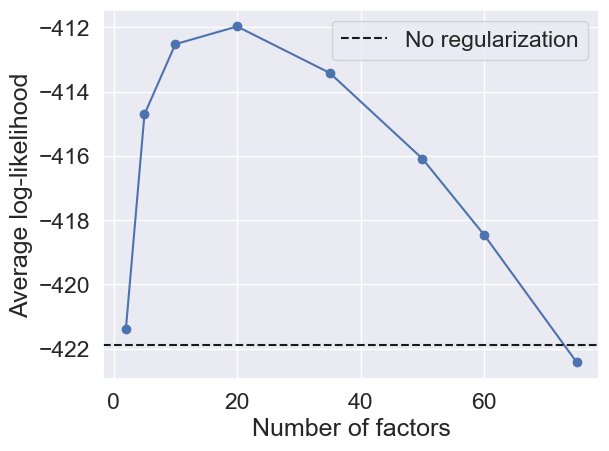}
\caption{Log-likelihood versus the number of factors, using a conventional 
factor model.}
\label{f-ll-vs-rank-trad-fact}
\end{figure}

\paragraph{Traditional factor model.}
Figure~\ref{f-ll-vs-rank-trad-fact} shows the log-likelihood versus the number
of factors $k$ for $k$ between~2 and~75.
A large increase in log-likelihood is attained with around 20
factors, as compared to using the full covariance matrix.
Thus using a traditional factor model and applying our 
covariance estimation method to the factor returns improves our overall 
covariance prediction.

\begin{figure}
\centering
\includegraphics[width=\textwidth]{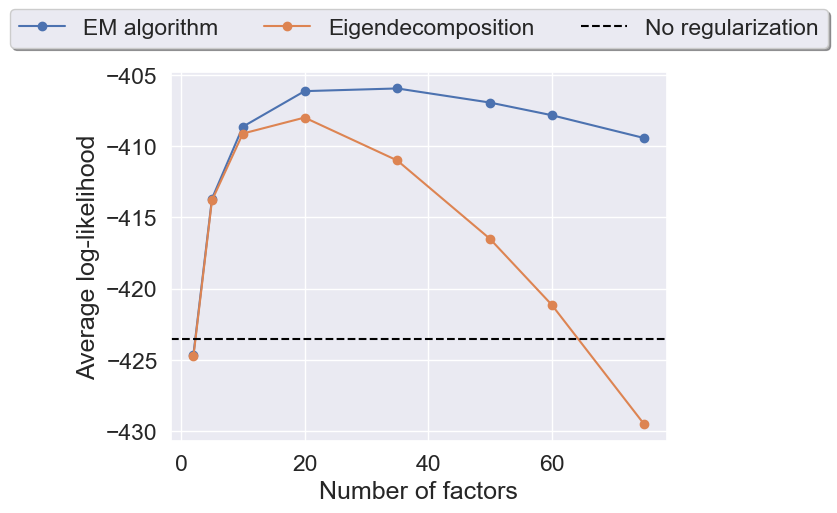}
\caption{Log-likelihood versus the number of factors, obtained by fitting
our covariance estimate with a factor model.}
\label{f-ll-vs-rank}
\end{figure}

\paragraph{Fitting a factor model to the covariance matrix.}
Figure~\ref{f-ll-vs-rank} shows the log-likelihood versus the number of factors
(\ie, the rank of the low-rank component) $k$ for various $k$ between~2 and~75,
using the eigendecomposition initialization and the EM algorithm.
We see that a rank of about $r=20$ seems optimal for this data set, and achieves
a noticeably higher log-likelihood than using the full-rank covariance.
Moreover, the EM algorithm does better than just computing the
eigendecomposition.

\begin{figure}
\centering
\subfigure[Risk.]{\includegraphics[height=0.33\textwidth]{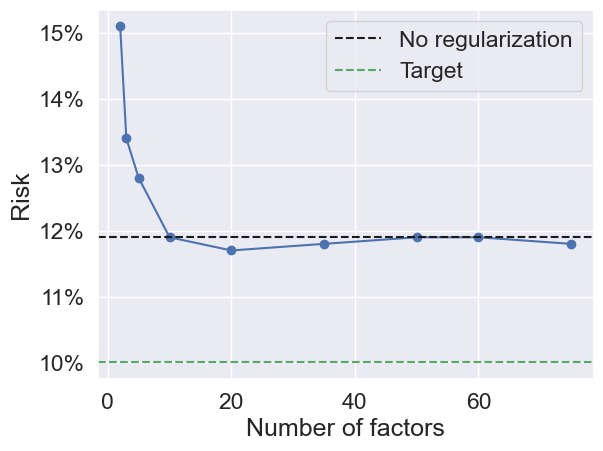}}
\\\subfigure[Sharpe ratio.]{\includegraphics[height=0.33\textwidth]{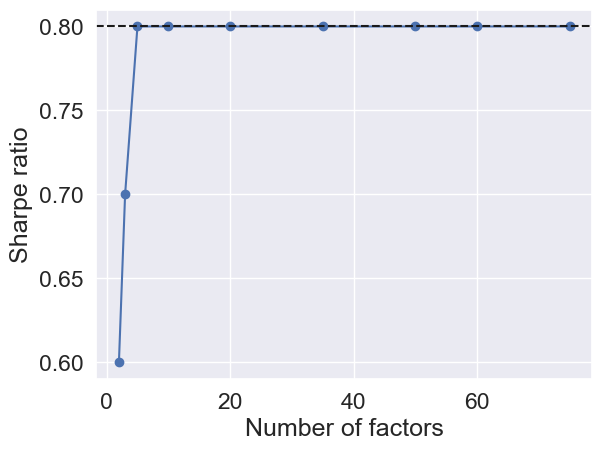}}
\\ \subfigure[Drawdown.]{\includegraphics[height=0.33\textwidth]{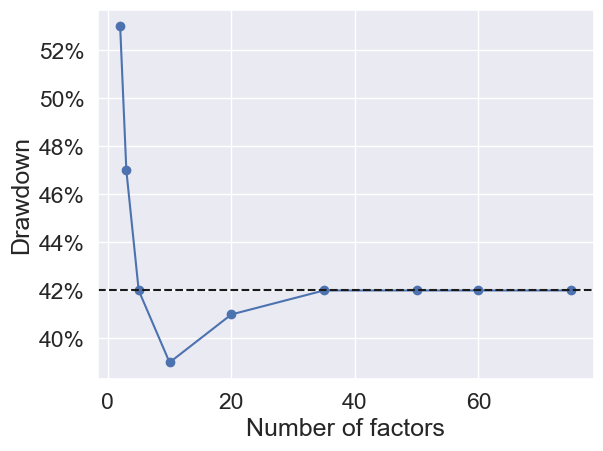}} 
\caption{Portfolio metrics for minimum variance portfolios constructed via factor models with various number of factors.}
\label{f-factor-portfolios}
\end{figure}

Figure~\ref{f-factor-portfolios} shows the portfolio metrics for the minimum
variance portfolios.  We can see that
with roughly 10 factors or more, the performance is essentially identical to 
that obtained using the full covariance matrix. 
For these experiments we observed no notable difference between the two factor model
fitting methods, \ie, the simple eigendecomposition based initialization and 
the more sophisticated EM method.
While using the factor model does not improve portfolio performance, it greatly
speeds up the computation of the portfolio optimization problems.

\chapter{Smooth covariance predictions}\label{c-smooth}
We address here a secondary objective for a covariance prediction
$\sigmahat_t$, which is that it vary smoothly across time.  Perhaps 
the main reason for desiring smoothness of the covariance estimate 
is that it can lead to reduced trading in portfolio construction
methods; it can also lead to improved portfolio performance, 
even without taking into account transaction costs. 

To some extent smoothness happens naturally, since whatever method is 
used to form $\sigmahat_t$ from $r_1, \ldots, r_{t-1}$ is likely to yield
a similar prediction $\sigmahat_{t+1}$ from $r_1, \ldots, r_t$.
It is also possible to further smooth the predictions over time, 
perhaps trading off some performance, \eg, in log-likelihood regret.

We have already mentioned that the weight optimization problem
\eqref{e-w-it-prob} can be modified to encourage smoothness of the weights over
time. 
We can also directly smooth the prediction $\sigmahat_t$, to get a smooth version
$\hat \Sigma_t^\text{sm}$.  A very simple approach is 
to let $\sigmahat_t^\text{sm}$ be a
EWMA of $\sigmahat_t$, with a half-life chosen as a trade-off between 
smoother predictions and performance.
This EWMA post-processing is equivalent to choosing
$\hat\Sigma_t^\text{sm}$ to minimize 
\[
\left\|\hat\Sigma_t^\text{sm}-\sigmahat_t\right\|_F^2+ 
\lambda \left\| \hat\Sigma_t^\text{sm}-
\hat\Sigma_{t-1}^\text{sm}\right\|_F^2,
\]
where $\lambda$ is a positive regularization parameter used to
control the trade-off between smoothness and performance,
or equivalently, the half-life of the EWMA post-processing.
Here the first term is a loss, and the second is a regularizer
that encourages smoothly varying covariance predictions.

We can create more sophisticated smoothing methods by changing the loss
or the regularizer in this optimization formulation of smoothing.
For example we can use the Kullback-Liebler (KL) divergence as a loss.
With regularizer $\lambda \|\hat \Sigma_t^\text{sm} - 
\hat \Sigma_{t-1}^\text{sm}\|_F$ (no square in this case), 
we obtain a piecewise constant prediction, which roughly speaking only
updates the prediction when needed.
This is a convex optimization problem which can be solved quickly and reliably
\citep{boyd2004convex}.

\section{Data and experimental setup}
We consider again the Fama-French factor returns from \S\ref{s-data}, over the
same time horizon. We use the CM-IEWMA covariance predictor with the same
parameters as in \S\ref{s-experiment-predictors}.
\paragraph{Smoothly varying covariance.}
In the first experiment we smooth the CM-IEWMA covariance estimates by applying
a EWMA, which corresponds to the $\lambda \left\| \hat\Sigma_t^\text{sm}-
\hat\Sigma_{t-1}^\text{sm}\right\|_F^2$ regularizer. For different EWMA
half-lives we attain different levels of smoothness.

\paragraph{Piecewise constant covariance.}
In the second experiment we smooth the CM-IEWMA covariance estimates by applying
the $\lambda \left\| \hat\Sigma_t^\text{sm}-
\hat\Sigma_{t-1}^\text{sm}\right\|_F$ regularizer. For different values of
$\lambda$ we attain piecewise constant covariance predictors with different update
frequencies.

\section{Empirical results}
\begin{figure}
\centering
\includegraphics[width=0.8\textwidth]{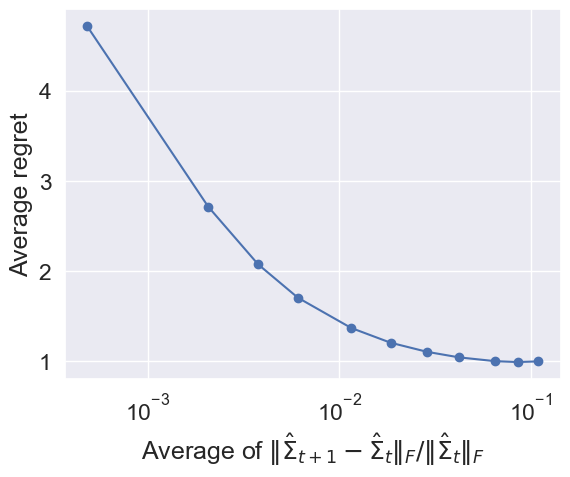}
\caption{Average regret versus smoothness when using EWMA smoothing of the covariance predictor.}
\label{f-ewma-smooth-covariance}
\end{figure}

\paragraph{Smoothly varying covariance.}
Figure~\ref{f-ewma-smooth-covariance} shows the regret versus smoothness for
various levels of smoothness.
As seen, we can reduce the smoothness by a factor of (roughly) four without
losing much performance in terms of regret. This can obviously be useful in
practice since a smoother covariance estimate would, for example, reduce
trading.

\begin{figure}
\centering
\includegraphics[width=0.8\textwidth]{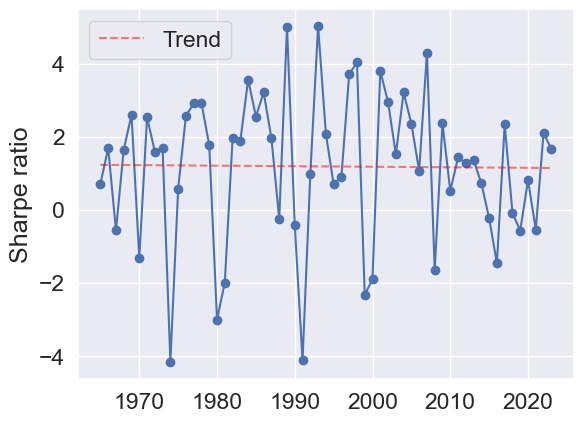}
\caption{Yearly annualized Sharpe ratios for the minimum variance portfolio when smoothing the CM-IEMA covariance predictor with a 250-day half-life EWMA.}
\label{f-ewma-smooth-yearly-sr}
\end{figure}

Table~\ref{t-l2reg} shows the portfolio metrics for various values of $\lambda$
for the minimum variance portfolio with the same parameters as in
\S\ref{s-portfolio-experiments}; here the turnover is defined as the average of
$252\times\|w_{t+1}-w_{t}\|_1/\|w_{t}\|_1$ over all times $t$ in the evaluation
period.
\begin{table}
\centering
\caption[width=0.1\textwidth]{Portfolio metrics for various EWMA half-lives used for smoothing the covariance. Half-life $0$ means no smoothing.}
\label{t-l2reg}
{\small
\begin{tabular}{lccccc}
\toprule
{Half-life/days} & {Return/\%} & {Risk/\%} & {Sharpe} & {Drawdown/\%} & {Turnover/\%} \\
\midrule
0 & 1.2 & 2.1 & 0.5 & 21 & $1855$ \\
10 & 1.4 & 2.1 & 0.7 & 16 & $310$\\
100 & 1.8 & 2.1 & 0.9 & 15 & $56$ \\
250 & 2.1 & 2.1 & 1.0 & \textbf{13} & $30$ \\
5000 & 2.9 & 2.6 & \textbf{1.1} & 21  & $9$ \\
\bottomrule
\end{tabular}}
\end{table}
Interestingly, the right amount of smoothing not only reduces turnover, 
but also improves portfolio performance in terms of Sharpe ratio and
drawdown, while keeping the desired volatility level. Too much smoothing, however,
leads to reduced portfolio performance. Figure~\ref{f-ewma-smooth-yearly-sr}
shows the yearly annualized Sharpe ratios for $\lambda=10^{-4}$, indicating a
stable performance over time.

\begin{figure}
\centering
\subfigure[No smoothing.]{\includegraphics[height=0.33\textwidth]{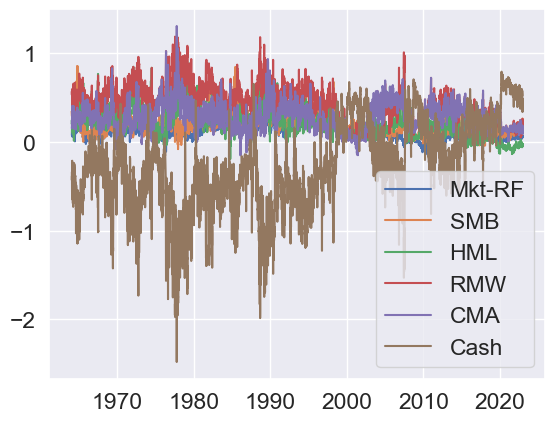}}
\\\subfigure[Half-life of 250 days.]{\includegraphics[height=0.33\textwidth]{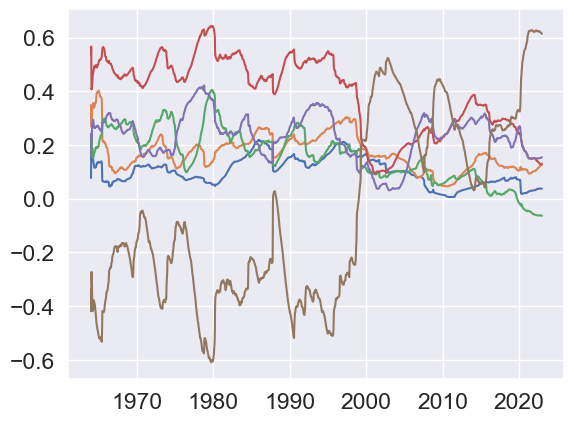}}
\\ \subfigure[Half-life of 5000 days.]{\includegraphics[height=0.33\textwidth]{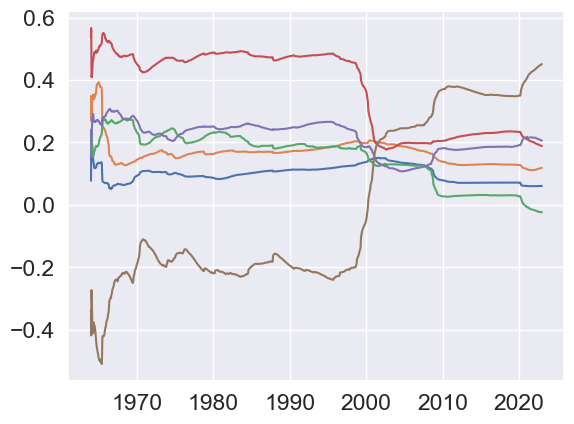}} 
\caption{Portfolio weights for three different regularization parameters $\lambda$.}
\label{f-L2smooth-port-weights}
\end{figure}

Figure~\ref{f-L2smooth-port-weights} shows the portfolio weights for three
different EWMA half-lives.
As seen, EWMA smoothing leads to smoothly varying portfolio weights, while the
weights vary significantly when no smoothing is applied.

\begin{figure}
\centering
\includegraphics[width=0.8\textwidth]{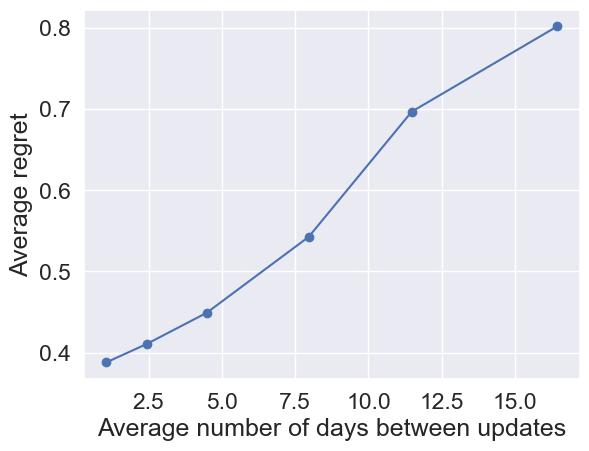}
\caption{Average regret versus time between covariance updates.}
\label{f-l1-smooth-covariance}
\end{figure}

\paragraph{Piecewise constant covariance.}
Figure~\ref{f-l1-smooth-covariance} shows the regret versus the update frequency
of the covariance estimate.
There is a clear trade-off between the regret and update frequency.
Roughly speaking, we could update the covariance matrix weekly without losing much in
terms of regret.

\begin{figure}
\centering
\includegraphics[width=0.8\textwidth]{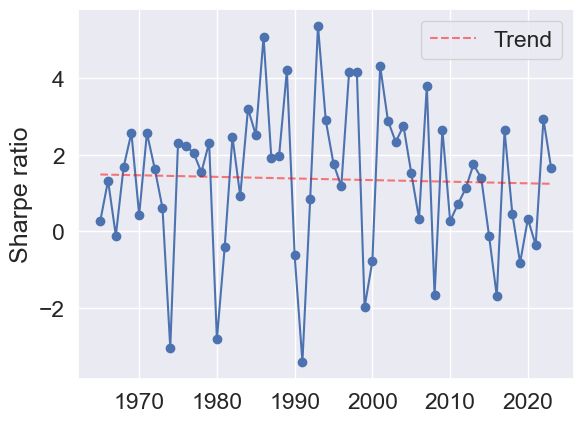}
\caption{Yearly annualized Sharpe ratios for the minimum variance portfolio with
a piecewise constant CM-IEMA covariance predictor using $\lambda=10^{-4}$.}
\label{f-l1-smooth-yearly-sr}
\end{figure}

As mentioned, a piecewise constant predictor can be desirable in practice, since
it encourages us not updating the portfolio weights, which in turn reduces
trading costs.
Table~\ref{t-l1reg} shows the portfolio metrics for various values of $\lambda$
for the minimum variance portfolio from \S\ref{s-portfolio-experiments}.
\begin{table}
\centering
\caption[width=0.1\textwidth]{Portfolio metrics for various regularization parameters $\lambda$. $\lambda=0$ means no smoothing.}
\label{t-l1reg}
{\small
\begin{tabular}{lccccc}
    \toprule
{$\lambda$} & {Return/\%} & {Risk/\%} & {Sharpe} & {Drawdown/\%} & {Turnover/\%} \\
\midrule
0  & 1.2 & 2.1 & 0.5 & 21 & $1855$ \\
$5\times 10^{-5}$ & 1.9 & 2.0 & 1.0 & 14 & $1190$ \\
$10^{-4}$ & 2.4 & 1.9 & \textbf{1.3} & \textbf{9} & $112$ \\
$10^{-3}$ & 2.6 & 2.1 & 1.2 & 17 & $7$ \\
$7.5\times 10^{-3}$ & 3.0 & 4.8 & 0.6 & 31 & $0$ \\
\bottomrule
\end{tabular}}
\end{table}
As seen, smoothing can significantly reduce turnover, and
interestingly improve the Sharpe ratio and drawdown noticeably while
maintaining the correct risk level. Figure~\ref{f-l1-smooth-yearly-sr}
shows the yearly annualized Sharpe ratios for $\lambda=10^{-4}$. The performance
is relatively stable over time, with a small downward trend.

\begin{figure}
\centering
\subfigure[$\lambda=0$.]{\includegraphics[height=0.33\textwidth]{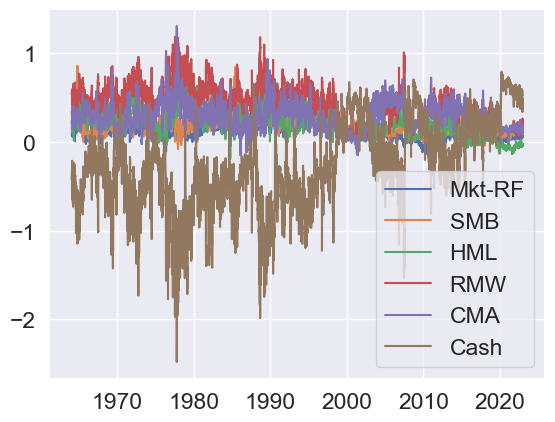}}
\\\subfigure[$\lambda=10^{-3}$.]{\includegraphics[height=0.33\textwidth]{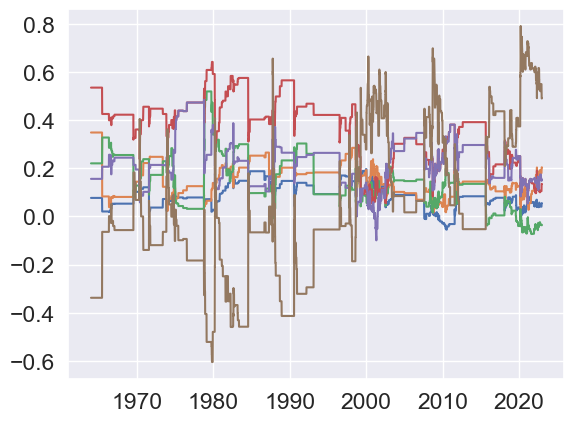}}
\\ \subfigure[$\lambda=10^{-4}$.]{\includegraphics[height=0.33\textwidth]{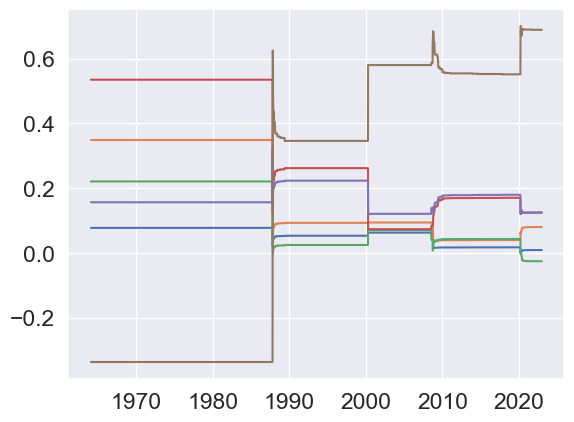}} 
\caption{Portfolio weights for three different regularization parameters $\lambda$.}
\label{f-L1smooth-port-weights}
\end{figure}

To illustrate the impact of smoothing we show the portfolio weights for three
different values of $\lambda$ in figure~\ref{f-L1smooth-port-weights}.
Without smoothing the portfolio weights are updated significantly every
day. For $\lambda=10^{-4}$ the weights are updated around once or twice a month.
Finally, for $\lambda=10^{-5}$ the weights are updated on average every half a
year, with only four big weight updates over the whole trading period.
Interestingly, the weight updates for $\lambda=10^{-5}$ correspond precisely in time
to the volatile regime around 1980, the 2000 dot-com bubble, the 2008 financial
crisis, and the 2020 pandemic. In short, we can conclude from
table~\ref{t-l1reg} and figure~\ref{f-L1smooth-port-weights} that smoothing can lead to
less trading and improve the portfolio performance. 

Finally, we note that there is some deviation between the regret metric and
portfolio performance. As seen from figure~\ref{f-l1-smooth-covariance} regret
increases as we update the covariance matrix less than every other
week. However, as seen from table~\ref{t-l1reg} and figure~\ref{f-L1smooth-port-weights},
portfolio performance can improve notably when updating the covariance
matrix only every few months, or even years.

\chapter{Simulating returns}\label{c-generative}

Our model can be used to simulate future returns, when seeded by past 
realized ones. To do this, we start with realized returns for periods 
$1, \ldots, t-1$, and compute $\sigmahat_t$ using our method.  
Then we generate or sample $r_t^\text{sim}$
from $\mathcal N(0,\sigmahat_t)$.  We then find
$\sigmahat_{t+1}$ using the returns $r_1, \ldots, r_{t-1}, r^\text{sim}_t$.
We generate $r_{t+1}^\text{sim}$ by sampling from $\mathcal N(0,\sigmahat_{t+1})$.
This continues.

This simple method generates realistic return data in the short term. 
Of course, it does not include shocks or rapid changes in the return statistics
that we would see in real data,
but the generative return method has several practical applications. To mention 
just one, we can simulate 100 (say) different realizations over the next quarter (say),
and use these to compute 100 performance metrics for our portfolio. This gives us
a distribution of the performance metric that we might see over the next
quarter.

\section{Data and experimental setup}
To illustrate the generative return method, we consider the five Fama-French factor
returns from \S\ref{s-data}. Using the same setup as in
\S\ref{s-experiment-predictors} we compute CM-IEMWA covariance estimates, using
data from January 1st 2011 to December 31 2013, \ie, over a three-year period.  Returns are then generated for
100 days, using the generative mode described above. 

\section{Empirical results}
\begin{figure}
\centering
\subfigure[Observed returns.]{\includegraphics[width=0.4\textwidth]{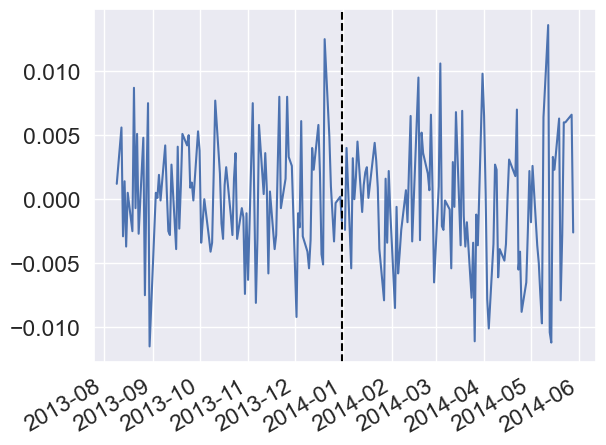}}
\\\subfigure[Obeserved returns (left) and simulated returns (right).]{\includegraphics[width=0.4\textwidth]{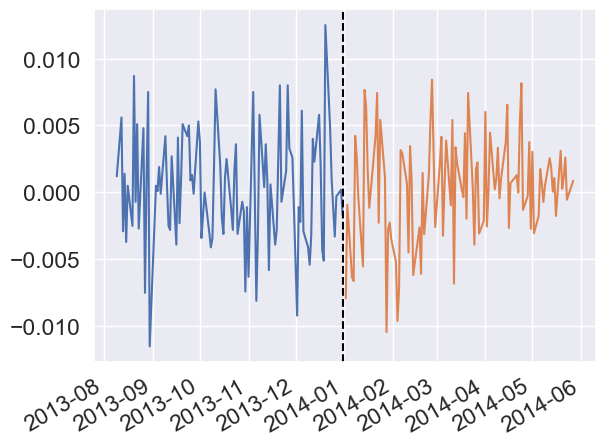}}
\\\subfigure[Observed returns (left) and simulated returns (right).]{\includegraphics[width=0.4\textwidth]{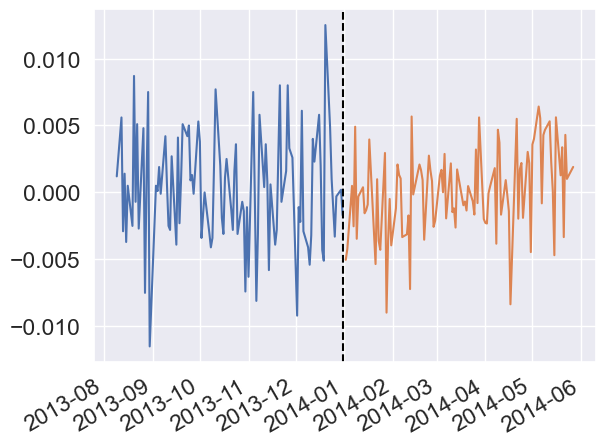}}
\caption{Observed and simulated SMB factor returns for two different seeds. The vertical line separates the in-sample (observed returns) and out-of-sample (simulated returns) periods.}
\label{f-SMB_true_simulated}
\end{figure}

We illustrate the results by looking at the SMB factor, \ie, we look at the
marginal distribution of this factor. Figure~\ref{f-SMB_true_simulated} shows
the true SMB factor returns and the simulated returns for two different random
number generator seeds.
As seen, we attain
realistic returns that could be used to generate scenarios for downstream
portfolio optimization tasks, for example.

\chapter{Conclusions}
We have introduced a simple method for predicting covariance matrices of
financial returns.  Our method combines well known ideas such as EWMA, first
estimating volatilities and then correlations, and dynamically combining
multiple predictions. The method relies on solving a small convex optimization
problem (to find the weights used in the combining), 
which is extremely fast and reliable. The
proposed predictor requires little or no tuning or fitting, is interpretable,
and produces results better than the popular EWMA estimate, and comparable to
MGARCH. Given its interpretability, light weight, and good practical
performance, we see it as a practical choice for many applications that require
predictions of the covariance of financial returns.

\begin{acknowledgements}
Stephen Boyd acknowledges many helpful in-depth discussions with his colleagues 
Trevor Hastie, Rob Tibshirani, Emmanuel Cand\a`{e}s, Mykel Kochenderfer, 
Kunal Menda, Misha Van Beek, Ron Kahn, and Gabriel Maher.
The authors thank Ron Kahn and Philipp Schiele for detailed comments and 
suggestions on early drafts.
The method of fitting a factor model to a Gaussian described 
in~\S\ref{s-factor-fitting} was suggested and derived by Emmanuel Cand\a`{e}s.
The authors are indebted to two anonymous reviewers for their detailed and 
helpful comments and suggestions.

\end{acknowledgements}

\backmatter  

\printbibliography

\end{document}